\documentclass[twocolumn,showpacs,amsmath,amssymb,amsthm,amsfonts,aps,pr]{revtex4} 
\usepackage{natbib}
\usepackage{graphicx}
\DeclareGraphicsExtensions{.jpg,.jpeg,.pdf,.png,.mps,.eps,.ps,.gif}
\usepackage{color}

\newcommand{\bminil}[1]{\begin{minipage}[l]{#1 \textwidth}}
\newcommand{\bminir}[1]{\begin{minipage}[r]{#1 \textwidth}}
\newcommand{\bminic}[1]{\begin{minipage}[c]{#1 \textwidth}}
\newcommand{\emini}{\end{minipage}}
\newcommand{\EQL}{\begin{equation}\label}
\newcommand{\EQ}{\begin{equation}}
\newcommand{\EN}{\end{equation}}
\newcommand{\BFG}{\begin{figure}}
\newcommand{\EFG}{\end{figure}}
\newcommand{\ITM}{\begin{itemize}}
\newcommand{\ITN}{\end{itemize}}
\newcommand{\ENM}{\begin{enumerate}}
\newcommand{\EEN}{\end{enumerate}}
\newcommand{\BEA}{\[\begin{array}}
\newcommand{\EEA}{\end{array}\]}
\newcommand{\EQAL}{\begin{eqnarray}\label}
\newcommand{\EQA}{\begin{eqnarray}}
\newcommand{\ENA}{\end{eqnarray}}

\newcommand{\dint}{\displaystyle \int}

\newcommand{\bs}{\mbox{\boldmath$s$}}

\newcommand{\bv}{\mbox{\boldmath$v$}}

\newcommand{\bx}{\mbox{\boldmath$x$}}

\newcommand{\ppt}{\frac{\partial}{\partial t}}

\newcommand{\half}{\mbox{$\frac{1}{2}$}}

\newcommand{\etall}{{\it et al.}}
\newcommand{\biband}{~and~}
\newcommand{\authone}[2]{#1~#2,}
\newcommand{\authtwo}[4]{#1~#2~and~#3~#4,}
\newcommand{\auththr}[6]{#1~#2,~#3~#4,~and~#5~#6,}
\newcommand{\authfour}[8]{#1~#2,~#3~#4,~#5~#6,~and~#7~#8,} 
\newcommand{\authmanytwo}[4]{#1~#2,~#3~#4,} 
\newcommand{\authmanythr}[6]{#1~#2,~#3~#4,~#5~#6,} 
\newcommand{\authmanyfour}[8]{#1~#2,~#3~#4,~#5~#6,~#7~#8,} 
\newcommand{\private}[2]{ (private communication).}

\newcommand{\yjour}[6]{``#6'' {#2} {\bf #3}, #4 (#1).}

\newcommand{\ybook}[3]{ {\em #2}. (#3, #1).}
 
\newcommand{\wwweb}[3]{#2, ``#3''(#1).}

\numberwithin{figure}{section}
\numberwithin{equation}{section}
\newcommand{\mycomment}[1]{} 
\newcommand{\figcomment}[1]{#1}
\newcommand{\bibcomment}[1]{}
\begin{document}
\title{Numerical generation of a vortex ring cascade in quantum turbulence}
\author{Robert M. Kerr}
\affiliation{Department of Mathematics, University of Warwick,
Coventry UK CV4 7AL}
\begin{abstract} A pair of perturbed anti-parallel vortices is
simulated using the three-dimensional Gross-Pitaevski equations.  
The simulations show decay of the vortex line length along with the 
depletion of local kinetic energy in a manner analogous to the decay 
of kinetic energy in the viscous, Navier-Stokes equations,
even though the governing equations are Hamiltonian and energy conserving.
The stages of development include vortex line stretching, 
conversion of kinetic energy into of the interaction component, 
the generation of vortex waves, multiple reconnections along
the waves leading to the creation of vortex rings, kinetic energy transfer 
to high wavenumbers, evidence for energy transfer into phonons and
finally the emission of the vortex rings. Some of this has been seen before,
but not in a single calculation, which demonstrates the existence 
of a mechanism for an energy cascade that does not rely on
statistical arguments. A four vortex example is given to demonstrate that
these steps might be general.  The wave generation and reconnection steps
all depend upon interactions between distinct vortices, unlike
vortex wave models where self-interactions along the vortices generate these effects.
A variety of additional cases are noted whose analysis is in progress.  So far,
none have generated the dynamics that vortex wave models require to generate a cascade.
Further cases are being pursued.
\end{abstract}
\maketitle
\section{Background}

Despite the absence of viscosity, experiments have repeatedly shown that
superfluids exhibit resistance and depletion of turbulent kinetic energy
in a manner similar to the effects of turbulence in a classical fluid. 
\cite{Smithetal93} made this quantitative, at relatively high temperatures,
by finding that the vortex line density decays at a rate consistent with the kinetic 
energy decay of idealized classical turbulence \citep{Kerrthesis81}.  
This result has since been confirmed at lower temperatures where
the possibility of coupling with the more classical, normal fluid component is
removed.  These experiments include $^3$He for $T<0.2T_c$ \citep{Bradleyetal05a},
where $T_c\leq 2.2$mK, and more recent measurements in $^4$He at $T\leq 0.5$K 
\citep{WalmsleyGolov08}. 
Besides the similarities to decay in classical fluids, superfluid spectral 
measurements yield classical $k^{-5/3}$ power laws, both for velocity
\citep{MaurerTabeling98} 
and for a signature of the vortex line density \citep{Rocheetal07}.  

Why would an inviscid, Hamiltonian system behave in this manner?
Unfortunately, our understanding of classical turbulent decay is so
poor that little insight is provided from that quarter.  We do not
even know if the equations we all believe underlie classical turbulence,
the Navier-Stokes equations, are well-posed.  What is observed in
high resolution numerical simulations of the Navier-Stokes equations is that 
the stretching of finite diameter vortices is crucial,
there is a $k^{-5/3}$ energy spectrum consistent with experiments and
ultimately the kinetic energy is converted into heat by viscous dissipation.
Most theories explaining the cascade are statistical and are not based
directly upon either the Navier-Stokes equations, or the vorticity equation.

In this paper, a series of steps towards a quantum turbulent state with
local kinetic energy depletion will be demonstrated in relatively modest 
three-dimensional numerical simulations
of the Gross-Piteavskii equations, the nonlinear Schr\"odinger equation 
most relevant to quantum fluids.
\EQL{eq:GP} \frac{1}{i}\ppt{\psi} = 0.5\nabla^2\psi + 0.5\psi(1-|\psi|^2) \EN
Unlike in classical fluids, the velocity does not appear in these equations. 
Only through the Madelung transformation does velocity appear 
and can defects in the wave function be interpreted as the equivalent of fluid vortices. 
Crucial differences with classical vortices are that quantum vortices are
infinitesimally thin and all have the same value of circulation.
Unlike the Navier-Stokes equations, these equations are a Hamiltonian system
and conserve their global energy and are regular for all time.  That is,
no higher-order derivative of the wave function can blow-up (have a
singularity) in a finite time. Despite these difference, reconnection
of vortices occurs in the inviscid Gross-Piteavskii equations, which in
classical systems is a purely viscous effect.

The primary initial state to be discussed consists of
two isolated, anti-parallel quantum vortices initially separated by several
core diameters (the healing length) with symmetric three-dimensional perturbations.  
It is found that these vortices transform themselves into a flow 
consistent with most of the observed properties of quantum turbulence.  In particular, 
the kinetic energy spectra increases in slope to be nearly -5/3 and kinetic energy is
removed from the interaction region, such that locally it appears as if
the energy is decaying.  It is found that wave dynamics and stretching
on the vortex lines play complementary roles in generating vortex reconnection.  
Neither by itself seems capable of generating the final state.

The most important new dynamics to be identified wil be a mechanism by which 
an energy cascade to small scales can be generated from the interaction of only
two vortices.  The required steps are:
\ITM
\item Vortex stretching generates a vacuum between the points where the vortices
are most strongly attracted.
\item There is an initial reconnection across this vacuum.
\item Curvature and torsion on the new vortex line created during the stretching
process generates a second reconnection event and the release of a vortex ring.
\item[*] During these phases, the vortex line length increases and
kinetic energy is converted into interaction energy..
\item Waves form further out along the original vortex lines, then deepen.
\item There are further reconnections leading to a state dominated by vortex rings.
\item A $k^{-5/3}$ kinetic energy spectrum forms at high wavenumbers in $k_y$.
\item The interaction energy is released from the original interaction region, 
either by breaking down into phonons or by the rings propagating to the boundaries.
\ITN

The paper will first present the equations, conservation properties, 
the numerical method and the initialization procedure.  
The discussion of the results will start by describing the evolution 
of the vortex topologies in physical space.  Next the development of
the high wavenumber spectra is discussed.  Finally, the mechanism for the depletion of
kinetic energy from the initial interaction region will be presented.
Several other cases are presented, including a four ring case that
support many aspects of the scenario suggested by the anti-parallel case.

\subsection{Simulating superfluids}

To simulate a nearly zero-temperature superfluid, the following properties 
must be satisfied.

\ITM\item Because a quantum fluid has no dissipation, its Hamiltonian, composed of 
gradient and interaction energies, 
should not decay in time except through additional physics or
through interactions with the boundaries.
\item A quantum fluid is compressible, with a complex relation
between the density and the pressure.
\item A quantum fluid is irrotational except along infinitely thin defects
in the wave function known as quantum vortices and these all have the
same quantum of circulation about their cores.  
\item[] The differences with classical vortices are:
\ITM
\item In a classical fluid, vorticity is distributed uniformly in space
and the circulation about vortex cores depends on the initial condition.
\item For an ideal classical fluid, these values of the circulation
are constant along Lagrangian trajectories.  But in a viscous fluid 
the circulation changes in time.
\ITN
\ITN

Simulations of the Gross-Piteavskii equations with good numerics and
adequate resolution should satisfy these properties.  However, these
equations must be discretized on a mesh and the numerical method must
be remain accurate even if the density goes identically to zero in
vortex cores.  Until recently, the numerics seemed incapable of achieving
these requirements.  For this reason, vortex methods are often applied.
However, vortex methods, whether Biot-Savart or the further reduced 
local induction approximation (LIA) have properties that might not
be consistent with the Gross-Piteavskii equations.  Foremost are:
\ITM\item In LIA vortices do not stretch, so the vortex line length cannot grow.
\item[] It will be demonstrated here, as have earlier Gross-Piteavskii calculations 
\citep{Leadbeateretal03,Berloff04}, that the line length grows
just prior to reconnection, indicating that vortices can stretch. And subsequent
to reconnection, the line length decreases.
\item One expects, and these calculations confirm, that reconnection can
lead to waves, twisting and further reconnections.  
\item[] However, both LIA and Biot-Savart calculations assume sharp reconnections, 
which at least in the orthogonal case \citep{Lipniacki00}, leads to 
exaggerated twisting and self-crossings that are not seen in Gross-Piteavskii calculations 
\citep{KoplikLevine93,AlamriYBarenghi08}.  
\ITN

This will now be mentioned further to explain why these artifacts 
that do not appear in simulations of the Gross-Piteavskii equations
nor in the experiments.

\subsection{Why not use a filament approximation?}

The first application of vortex methods to superfluids used
the local induction approximation (LIA) \citep{Schwarz78}.  In this approach, 
long-range Biot-Savart interactions are ignored, leaving only 
local dynamics along the vortex line, that is between neighboring points.
Along a single line these equations can be transformed into a 
1D nonlinear Schr\"odinger equation \citep{Hasimoto72}, which supports 
Kelvin waves and solitons.  Both phenomena are observed
in simulations of vortex lines, whether classical hydrodynamics,
Biot-Savart filament calculations, along GP vortices, as will be shown here,
and perhaps most dramatically the twists that go up and down tornadoes.

While LIA explains some dynamical features for an isolated filament, dynamics
due to interactions between vortices are excluded, including stretching and
reconnection.


\section{Equations, numerics and initial condition}

The equations simulated are a form of the standard higher dimensional 
focusing nonlinear Schr\"odinger equation for a complex wave function 
$\psi(\bx)$ with a cubic nonlinearity: The Gross-Pitaevskii equations \eqref{eq:GP}.
This is a hard-sphere approximation to the full equation of state:
\EQL{eq:GPfull} -i\hbar\ppt{\psi}(\bx) = (\hbar^2/2m)\nabla^2\psi(\bx) + \EN
\vspace{-6mm}
$$ \psi\int|\psi(\bx')|^2V(|\bx-\bx'|) d^3x' - E_v\psi(\bx) \,. $$
Following \cite{Berloff04}, the following choices are made:
$V(|\bx-\bx'|)=0.5\delta(\bx-\bx')$, the chemical potential
$E_v=0.5$ and $\hbar$ and $m$ are non-dimensionalized to be 1.  
While only an approximation for a superfluid, the hard-sphere approximation
is believed to be valid for a Bose-Einstein condensate
and is believed to qualitatively incorporate the most important physics 
for quantum turbulence in a very low temperature superfluid.  
Therefore, while idealized, it is physically relevant.

The outer
boundary condition for a BEC in a trap would have the wave function $\psi$ and
density $|\psi|^2$ go to zero at the boundaries, that is Dirichlet.  For
a superfluid in a container, it is better to use no-flux or Neumann boundary 
conditions, which is used here both as the outer boundary and as a way to
generate symmetric initial conditions about the point of interest, 
as described below.

Applying the usual Madelung transformation for quantum wave functions, if 
$\psi=\sqrt{\rho}\exp(i\theta)$ then the velocity is
\EQL{eq:Madelungv} \bv=\nabla\theta={\rm Im}\psi^\dag\nabla\psi/\rho\quad . \EN
These equations conserve mass:
\EQL{eq:GPmass} M = \int dV |\psi|^2 \EN 
and a Hamiltonian
\EQL{eq:GPHamiltonian} H = \half\int dV \left[ \nabla \psi \cdot \nabla \psi^\dag 
+ 0.5(1-|\psi|^2)^2 \right] \EN
where $\psi^\dag$ is the complex conjugate of $\psi$ and
\EQL{eq:GPenergies} K_{\nabla\psi}=\half|\nabla\psi|^2
\quad{\rm and}\quad E_I(\bx)=\half(1-|\psi|^2)^2  \EN
$K_{\nabla\psi}$ is the kinetic or gradient energy and $E_I$ is the interaction energy.

The advantage of considering only these two components of the energy is that
this decomposition also applies in Fourier space and by examining the 
interaction terms in detail in Fourier space, one can more fully understand
how a cascade might develop in that space.  This is developed further in the
section on spectral analysis and will be the subject of a subsequent paper.

\subsection{Numerical method}

The numerics is a standard semi-implicit spectral algorithm where the
nonlinear terms are calculated in physical space, then transformed to
Fourier space to calculate the linear terms.  In Fourier space,
the linear part of the complex equation is solved through integrating factors with
the Fourier transformed nonlinearity added as a 3rd-order Runge-Kutta explicit forcing. 
The domain is imposed by using no-stress cosine transforms in all
three directions. 

This scheme has been used successfully for another ideal equation,
the incompressible 3D Euler equations \cite{BustamanteKerr08} where
the timestep can be chosen using the Courant condition and the kinetic
energy is conserved to within machine accuracy.  Because the
velocity has no role in the integration of the GP equations and because
the nonlinearity is cubic, not quadratic, it is not possible to choose the
timestep in a similar way to conserve the total energy. 

Instead, a resolution dependent constant timestep is used that is chosen to be
small enough that the results converge in a manner consistent with the 
Runge-Kutta algorithm.  A measure of the accuracy
of the method is to track the Hamiltonian over time, which grows slightly
with time, roughly as $(\delta t)^3$.  Over the course of the highest
resolution calculation reported here, with 20,000 time-steps, 
it grew by a factor of 0.01. 

The boundary conditions serve two purposes.  Because
the initial condition has two symmetries, it incorporates those symmetries
into the numerics at the appropriate boundaries.  Secondly, on the other
boundaries the cosine transforms provide impenetrable, stress-free 
or free-slip boundaries, which are a reasonable approximation to the 
true boundary conditions in a superfluid.  If more realistic conditions
were required at the outer boundaries, extra terms could be added that
could extract energy from the system.  However, the results presented here demonstrate
that even for the ideal equation there are effects that can mimic dissipation 
without relying upon any additional dissipative terms.

\subsection{Initial conditions}

\begin{figure}
\figcomment{
\includegraphics[scale=.1]{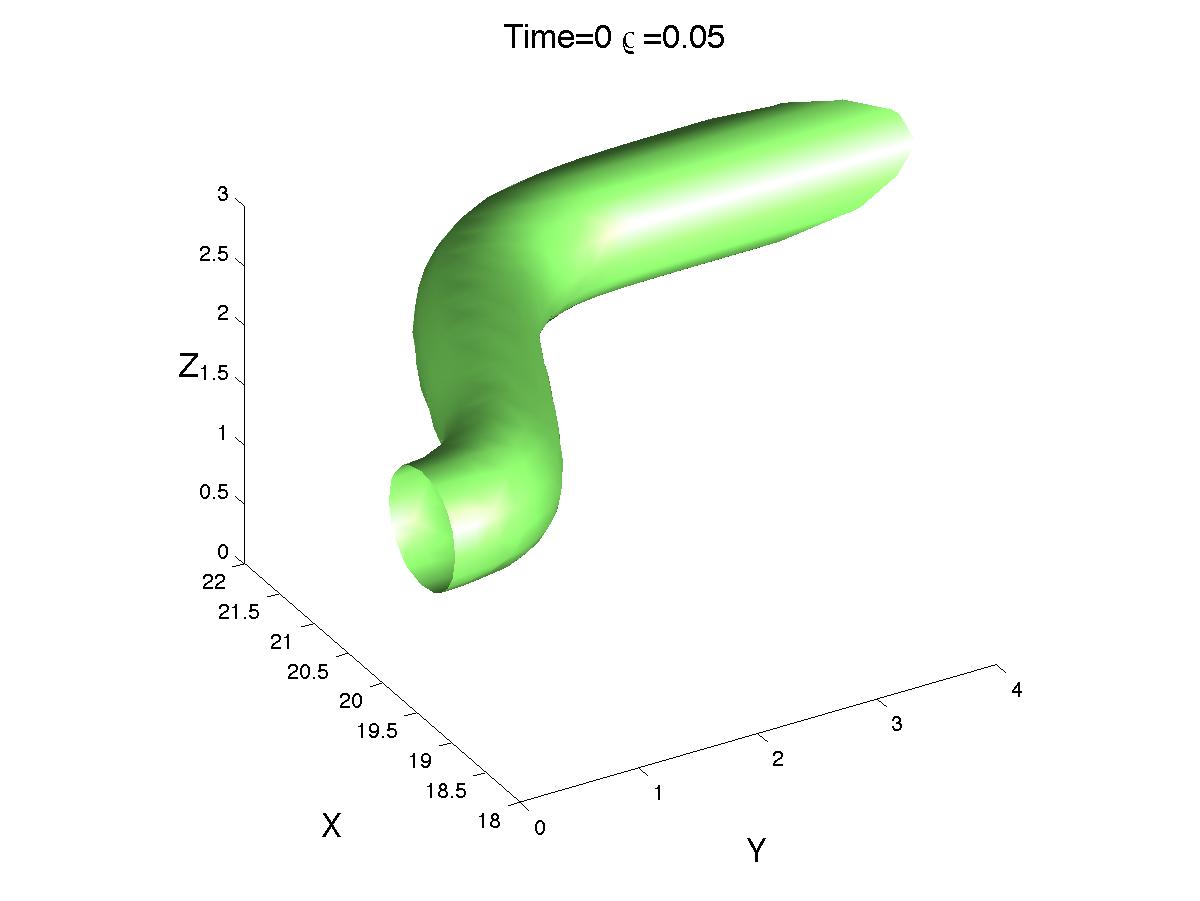}
\includegraphics[scale=.1]{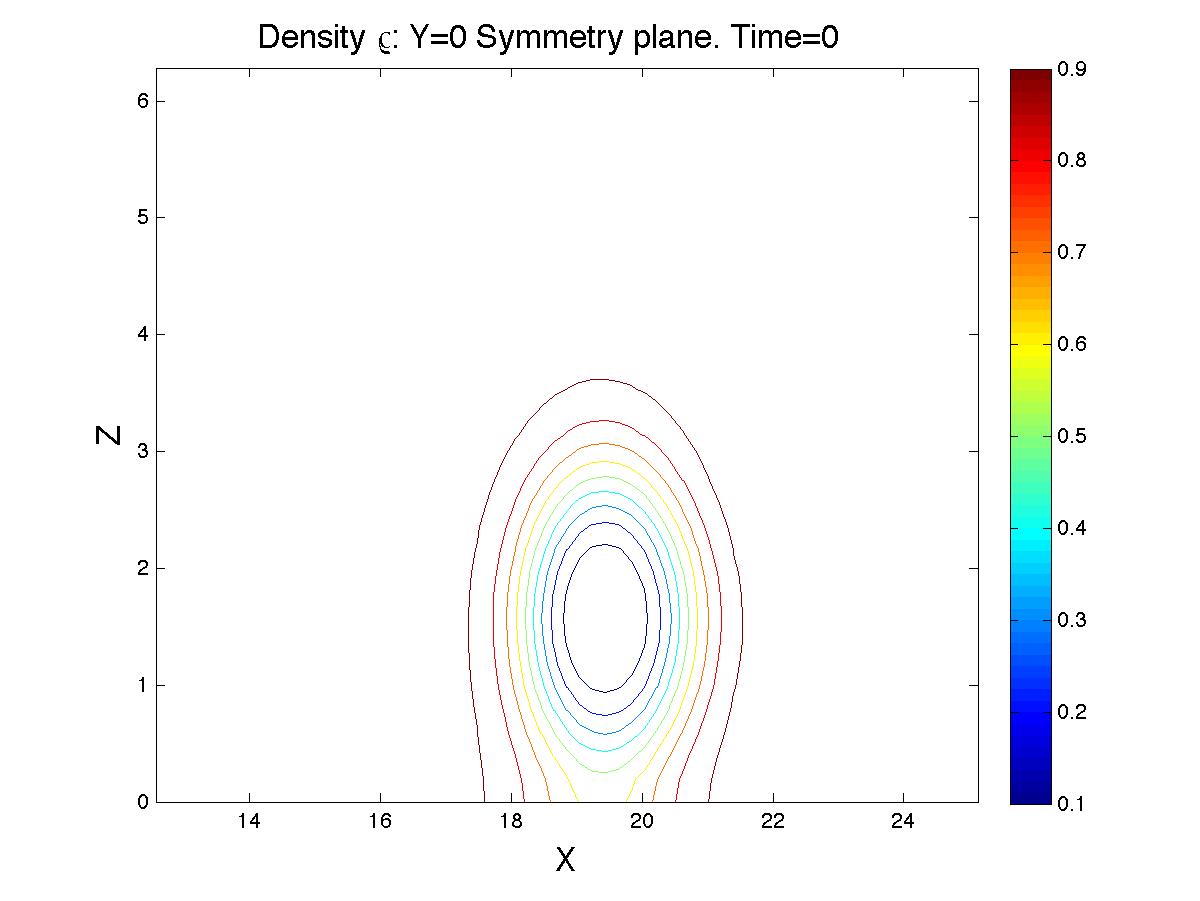}
}

\caption{Zooming in on the initial $T=0$, vortex. 
Left: Three-dimension isosurface with $\rho=.05$. 
Right: Density contours through the $y=0$ symmetry plane of maximum perturbation.
Recall that this is only one half of one of two mirrored vortices. 
A complete geometry for a similar initial condition appears in
\cite{BustamanteKerr08}.}
\label{fig:IC}
\figcomment{
\includegraphics[scale=.1]{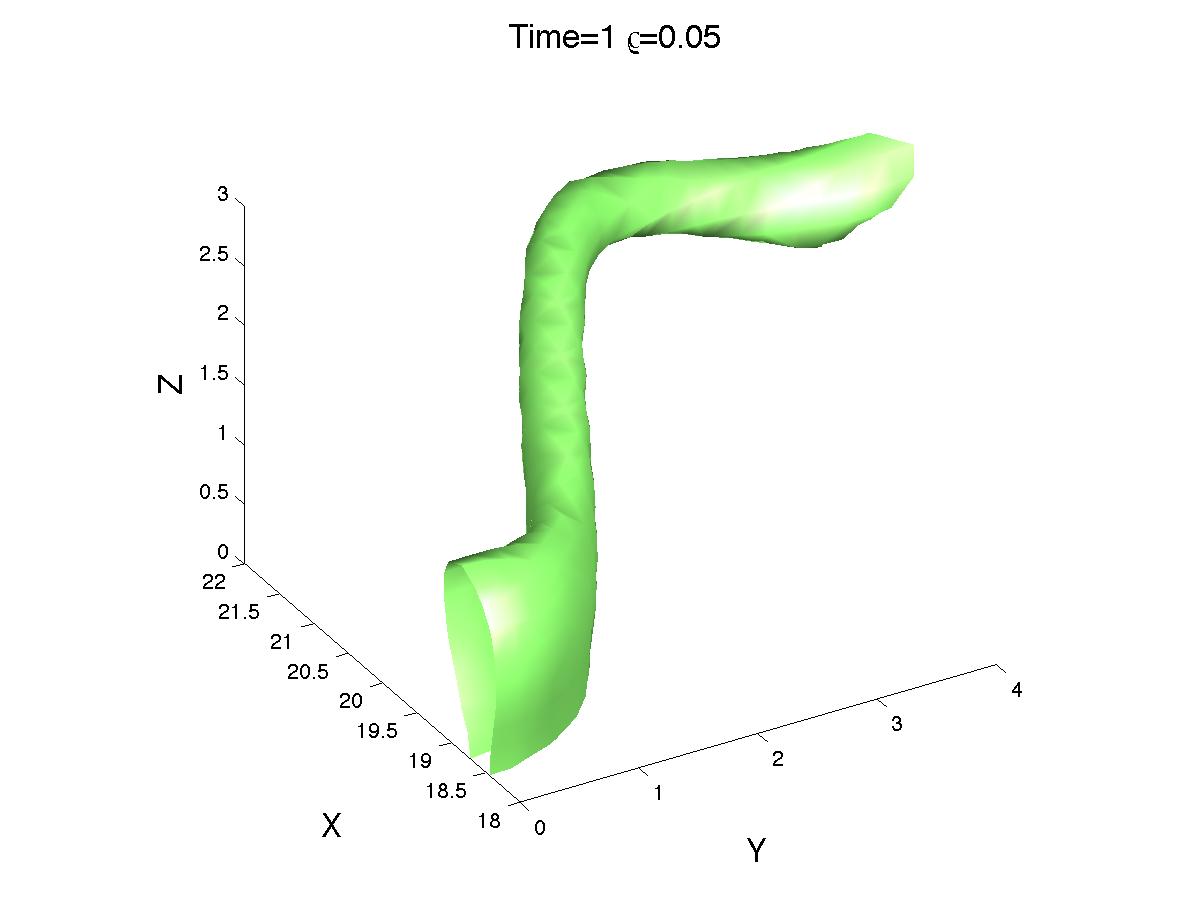}
\includegraphics[scale=.1]{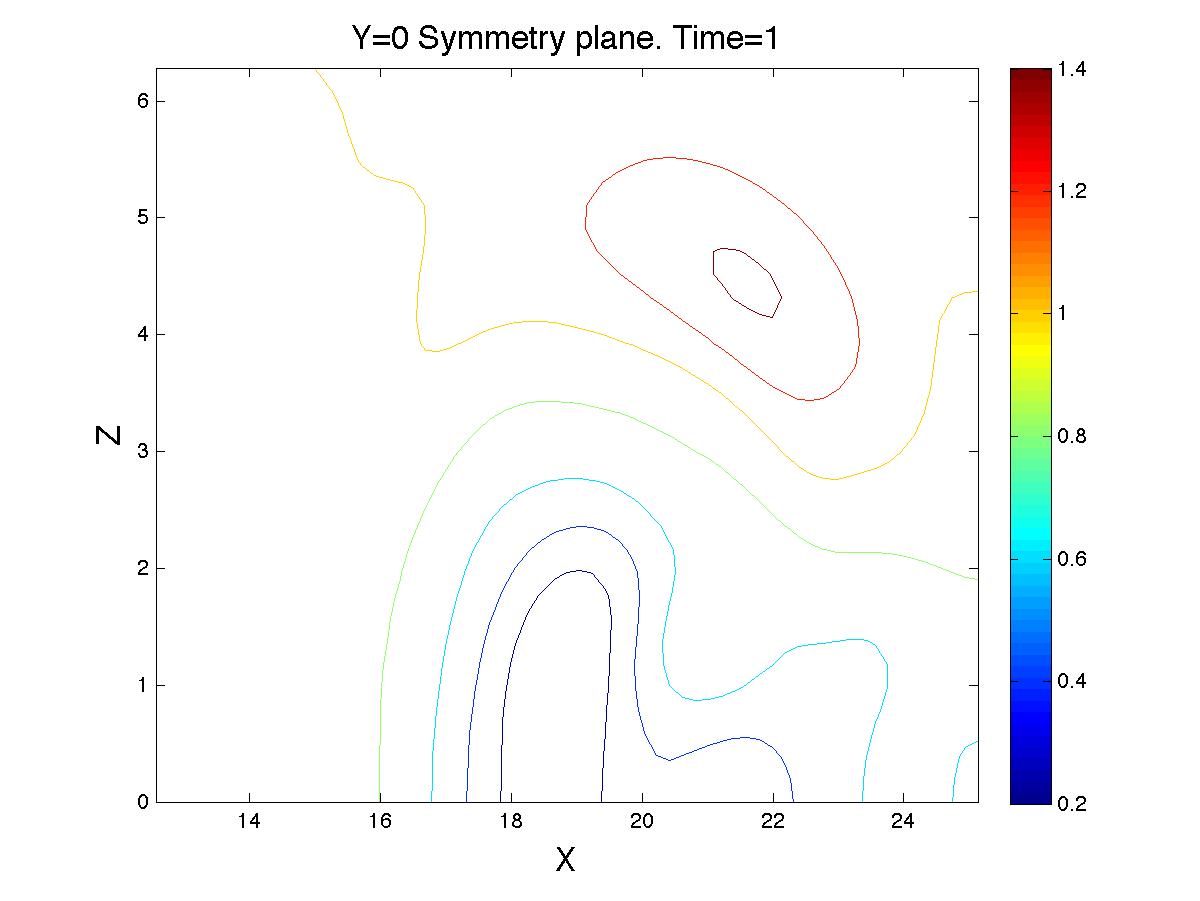}
}
\caption{Early evolved time, $T=1$. 
Left: Three-dimension isosurface with $\rho=.05$.  Right: $y=0$, $\rho$ contours.
First, as in a classical fluid, due to vortex stretching, the isosurface thins,
most evident for the segment near $y=z=2$.  Near $y=0$ at this time, the stretching has
pulled sufficient density away that an extended gap of $\rho\approx 0$
is opening up.  Reconnection will occur across this gap.}
\label{fig:kinkgapT1}
\end{figure}

\begin{figure} 
\figcomment{
\includegraphics[scale=.1]{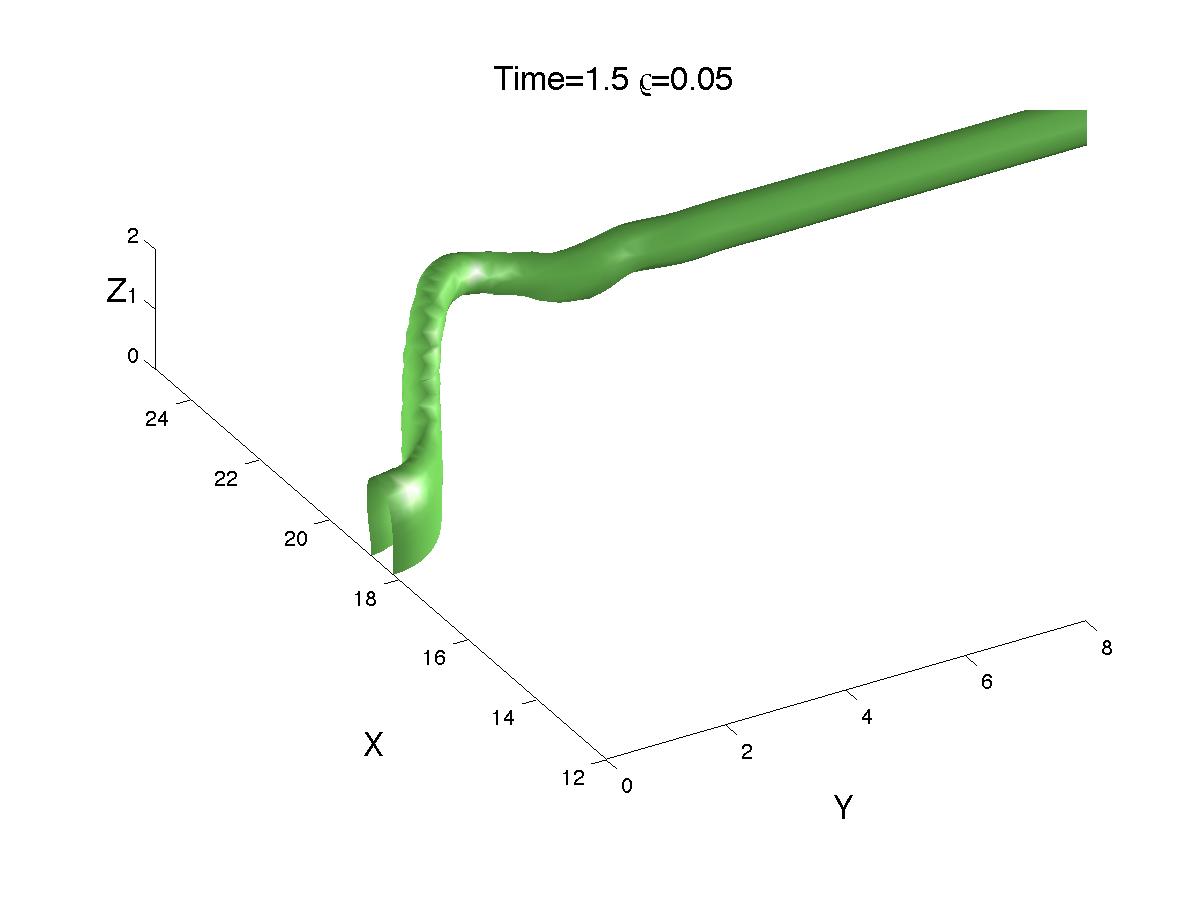}
\includegraphics[scale=.1]{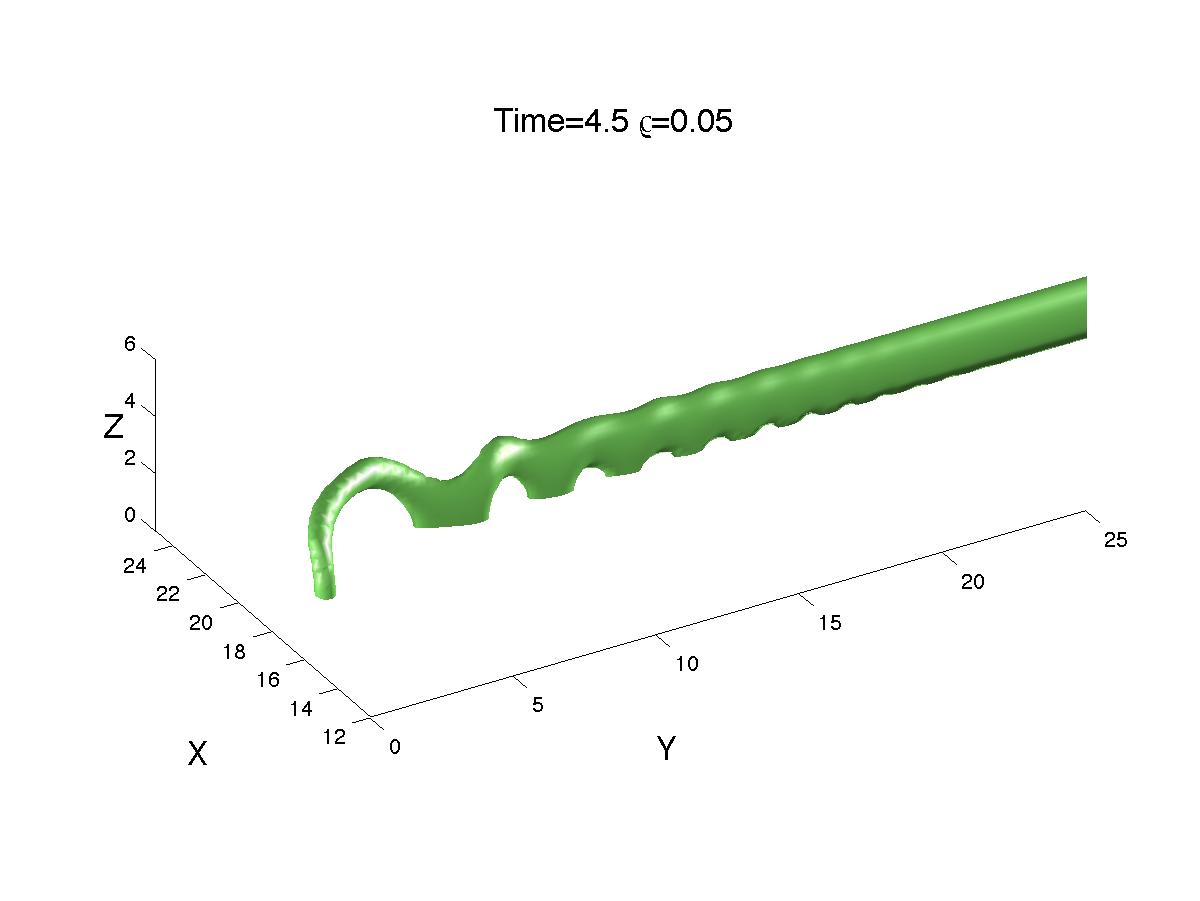}
}
\caption{Leading up to reconnection, a kink appears at the transition between
where the collapsing highly twisted vortex meets its straight continuation.
Out of this kink, waves propagate outward along the vortex.  
After reconnection, the waves don't simply propagate.  They deepen until
a second reconnection occurs and a vortex ring separates for the origin vortices.}
\label{fig:waves} 
\figcomment{
\includegraphics[scale=.1]{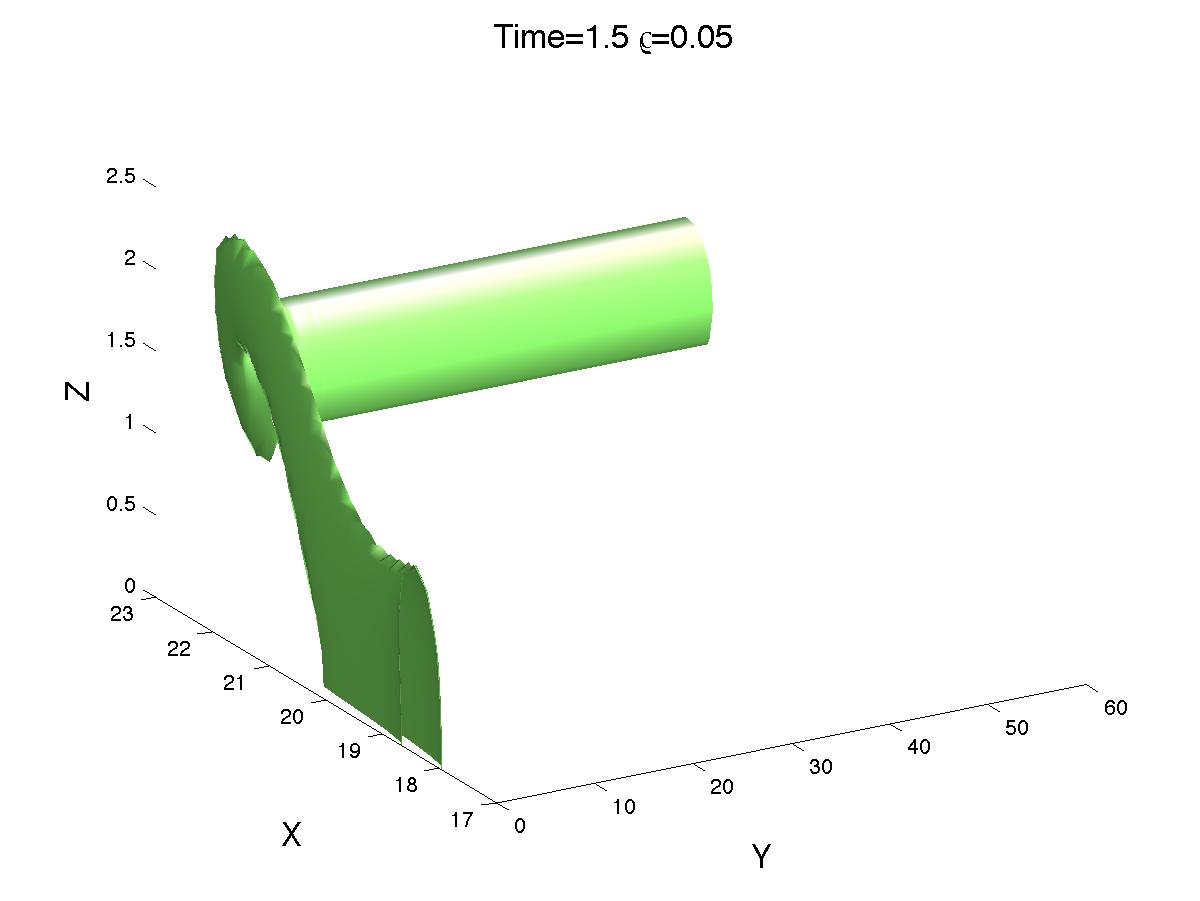}
\includegraphics[scale=.1]{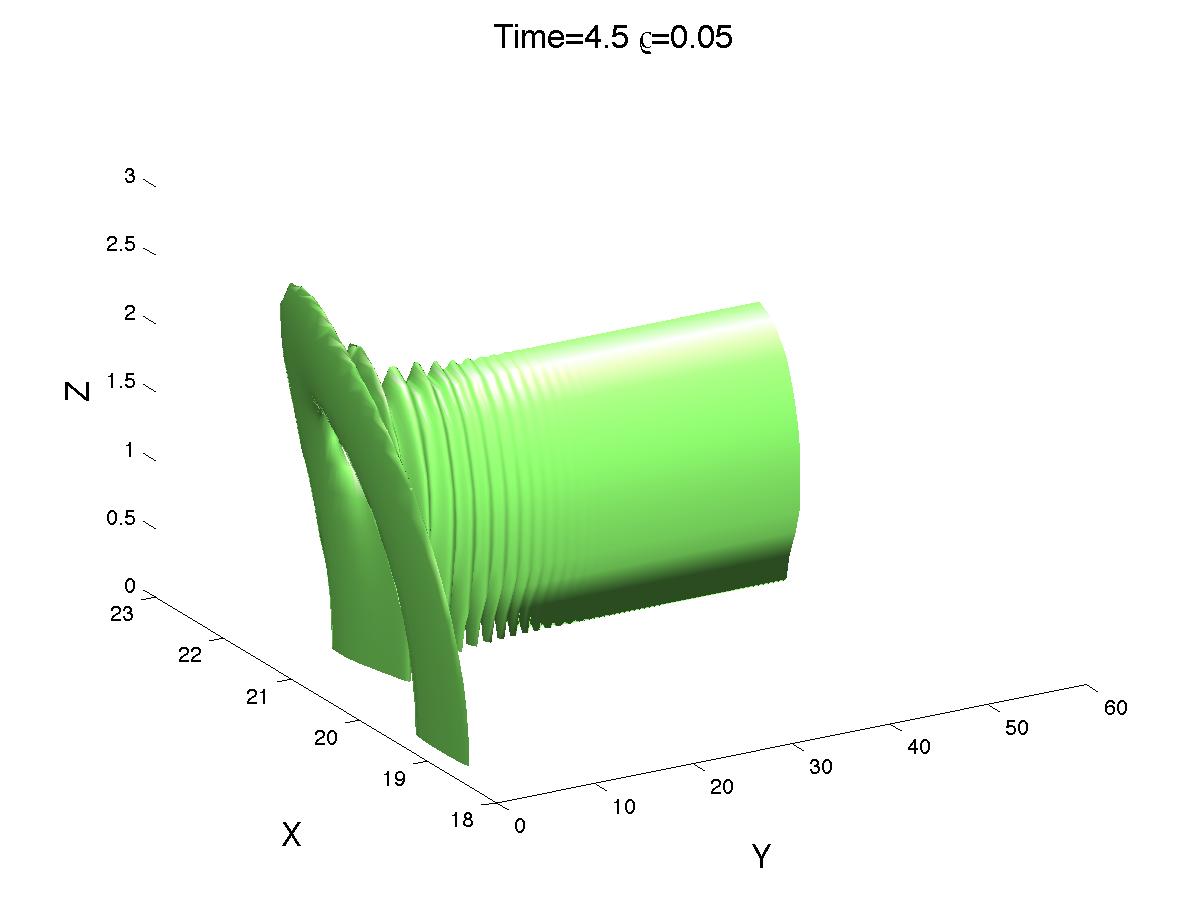}
}
\caption{By foreshortening the $y$ direction, the developing twisting can
be better illustrated.}
\label{fig:forewaves} 
\end{figure} 

The goal for initialization is to provide a smooth initial condition along 
a curved vortex.  Due to the symmetries imposed, one could initialize with
only one vortex.  However, it was soon realized that this resulted in
discontinuities of higher-order derivatives of the wave function at the
boundaries, and as soon as the cosine decomposition was imposed, waves
appeared throughout the domain.  
These discontinuities were removed by using multiple images
from outside the domain of the initial condition and a Fourier smoothing.
Earlier work \citep{KoplikLevine93} used three secondary images.  In
this work, 23 secondary images are used. 

Another feature that is required for a smooth initial condition is that
the density profile be both smooth and respects any expected analytic
balance between centrifugal forces and pressure about vortices.  The
most singular part of the initialization procedure is around the vortex cores, so to be
on the safe side the initialization in the vortex cores was chosen smoother
than might have been necessary.
The final step for obtaining a wave-free, smooth initialization
is to filter the highest wavenumber with a $\exp(-\alpha k)^4)$ filter.
Following experience gained initializing for the incompressible Euler equations
\cite{Kerr93}, the final filter should be used only to remove any final 
extraneous noise, and not be the primary means by which inconsistencies are removed.
A measure of how successful all the other steps are at generating 
a wave-free, smooth state is how small the filter is needed.  
$\alpha=.002$ when $\max(k_z)=5$ was found to be sufficient and in the worst test
case reduced the pre-filter Hamiltonian by 2\% and the density changed
by .02\%. 

The basic density profile about the singularities in the vortex cores
obeyed an analytic profile of the form 
\EQL{eq:initprofile} |\psi|=\sqrt{\rho}=r^m/\sqrt{r^m+a_0} \EN
While it is based upon a Pad\'e approximate to an ideal 2D quantum vortex \citep{Berloff04a},
which has $m=1$ and $a_0=2$, for these calculations $m=2$ and $a_0=2$
was chosen to ensure that the wave function around the vortex singularities
was as smooth as possible initially.  Once the calculation started, this
profile quickly adjusted to $m=1$ profile.

The initial profile function 
was applied approximately perpendicular to the trajectory of the
vortex lines, and not perpendicular to the $y$-axis
as was done for an Euler calculation \citep{BustamanteKerr08}.  
The reason for the change is that with the former procedure,
there were large ($>50\%$) variations in the density above the background of 
$\rho=1$ in the initial condition.  Using the new method
the maximum excess density peaks were reduced to $\rho<1.1$. 

Two other differences with the Euler simulations are the trajectory
of the vortex and the size of the domain in the vortical $y-$direction.
Both of these modifications were dictated by the formation of post-reconnection
Kelvin waves on the vortices. For
a smaller domain these would propagate to the $y$-domain wall and 
reflect back by the time only one vortex ring had separated off.
To see the multiple rings reported required that the domain wall be set back
further, and that there be no perturbations to the initial anti-parallel
trajectories beyond a short distance in $y$.  So rather than the
nested cosine trajectory used for Euler \citep{Kerr93}, the
new trajectory path is:
\EQL{eq:newtraj} \bs =\left(\delta_x\frac{2}{\cosh\bigl((y/\delta_y)^{1.8}\bigr)}-1,1,0\right) \EN
with $\delta_x=-1.6$ and $\delta_y=1.25$
in a $L_x\times L_y\times L_z = 8\pi\times 16\pi\times 4\pi$ domain on a
$128 \times 512 \times 64$ mesh. The power 1.8 on the normalized position $y/\delta_y$
was chosen so as to localize the perturbation near the $y=0$ symmetry plane.
This new trajectory assisted in minimizing the maximum excess density spots. 

\section{Results: Stretching, reconnection, waves, rings}

The first three figures are designed show the stages in development
before vortex rings begin to form.
Figure \ref{fig:IC} shows two early two isosurfaces of density for the 
initial condition chosen.  Fig. \ref{fig:kinkgapT1} 
shows the initial vortex stretching. And Fig. \ref{fig:waves}
shows the evolution from this initial condition to a state where the
vortex waves have started to reflect off the external wall.

\paragraph{Stretching} That vortices stretch in the Gross-Piteavski equations 
was first suggested by \citep{Noreetal97} using spectral properties.
Using a box counting method \cite{Leadbeateretal03} provided evidence for
both the growth and decay of vortex lines.
How a pair can stretch is demonstrated here by 
Fig. \ref{fig:kinkgapT1} and quantified in Fig. \ref{fig:linel}.
For a superfluid, vortex stretching changes the density between the vortex
cores.  First drawing denser fluid in and making the region of 
$\rho\approx 0$ thinner, then as the two vortices approach
one another, creating a vacuum.  

\paragraph{Vacuum and reconnection} The creation of the pre-reconnection vacuum 
starts on the $x-z$ symmetry plane 
then extends to the $x-y$ dividing plane at $t=1$.  It is through this vacuum that 
reconnection occurs.  The creation of the vacuum results in
a transformation of kinetic energy into interaction energy.

\paragraph{Waves}
Following reconnection and the creation of new vortices with sharp curvature, 
Kelvin waves are generated along the new vortices, as shown in Fig. \ref{fig:waves}.
Similar waves were seen using the local induction approximation following 
a wall-induced reconnection event \citep{Schwarz85}
and have been explained using LIA \citep{Svis95,Lipniacki00}. The
self-crossing of the vortex lines noted there bears some resemblance to the 
deepening of the vortex lines following reconnection reported here.

\subsection{Generation of rings}

The vortex waves continue to deepen until they reconnect again
across the x-y dividing plane in Fig. \ref{fig:waves}.  This is best
illustrated by the $T=4.5$ isosurface in Fig. \ref{fig:waves}.  Closer
examination reveals a series of these kinks is developing for $y\geq 5$
as the vortex waves become a series of zig-zags.  
A second reconnection at the kink near $y=5$ results in a ring separating 
from the paired filaments.

After the second reconnnection and separation of the first vortex ring, a succession of
reconnections quickly occur between the wavy vortices, that is between
the illustrated vortex and its mirror across the $x-y$ dividing plane,
which lead multiple rings.
Fig. \ref{fig:rings} shows the isosurfaces and $x-y$ dividing plane
cross-sections of density to show
where these subsequent reconnections occur. 

Note that each successive ring that forms has a smaller
radius than the previous ring. Because the quantum circulation about the vortex
cores of each vortex ring is identical, this implies that the
propagation velocity $\mathbf V\sim \Gamma/R$ of each ring increases. 
and the separation between the rings
increases. This implies they will not interfere with each other
and can therefore freely leave the local system.

Why are the successive rings each smaller than the previous one?
Could it be an artifact of the particular initial condition of two
nearly anti-parallel vortex lines?  The four vortex case mentioned below
demonstrates a state similar to this when the tangle is formed.
Higher resolution and further analysis will be necessary to determine which 
secondary properties associated with the formation of multiple rings, 
exist in this calculation as well.

Where, in physical space, could a cascade be forming that
the spectral analysis reported next finds?  Since the rings separate with
time, it cannot be forming from subsequent ring-ring interactions and
further reconnection.  Instead, the ongoing analysis suggests that each spectral
cascade step is associated with the formation of each new ring.  That is,
each new and smaller ring pulls energy into its smaller length scale
and corresponding higher wavenumber.

\BFG[!]\label{fig:rings} 
\figcomment{
\includegraphics[scale=.1]{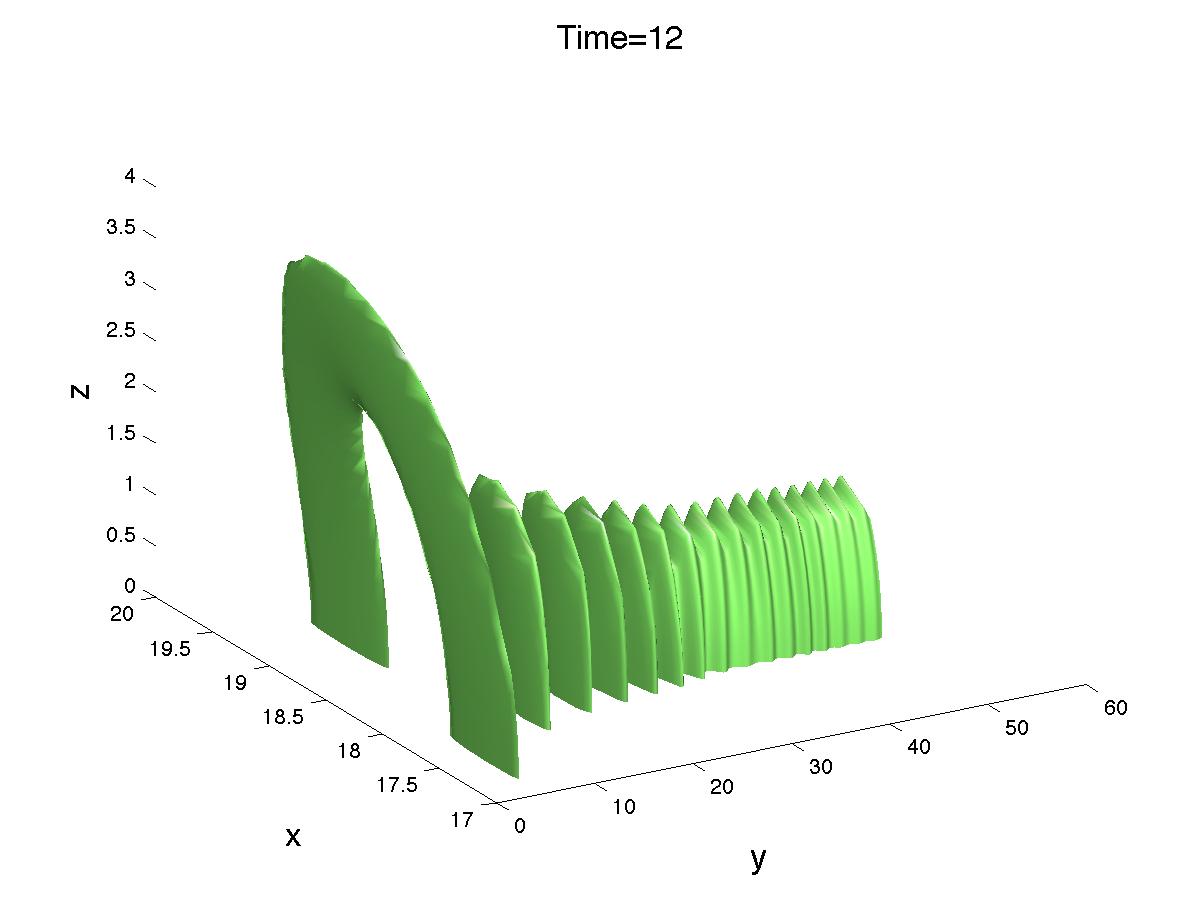}
\includegraphics[scale=.1]{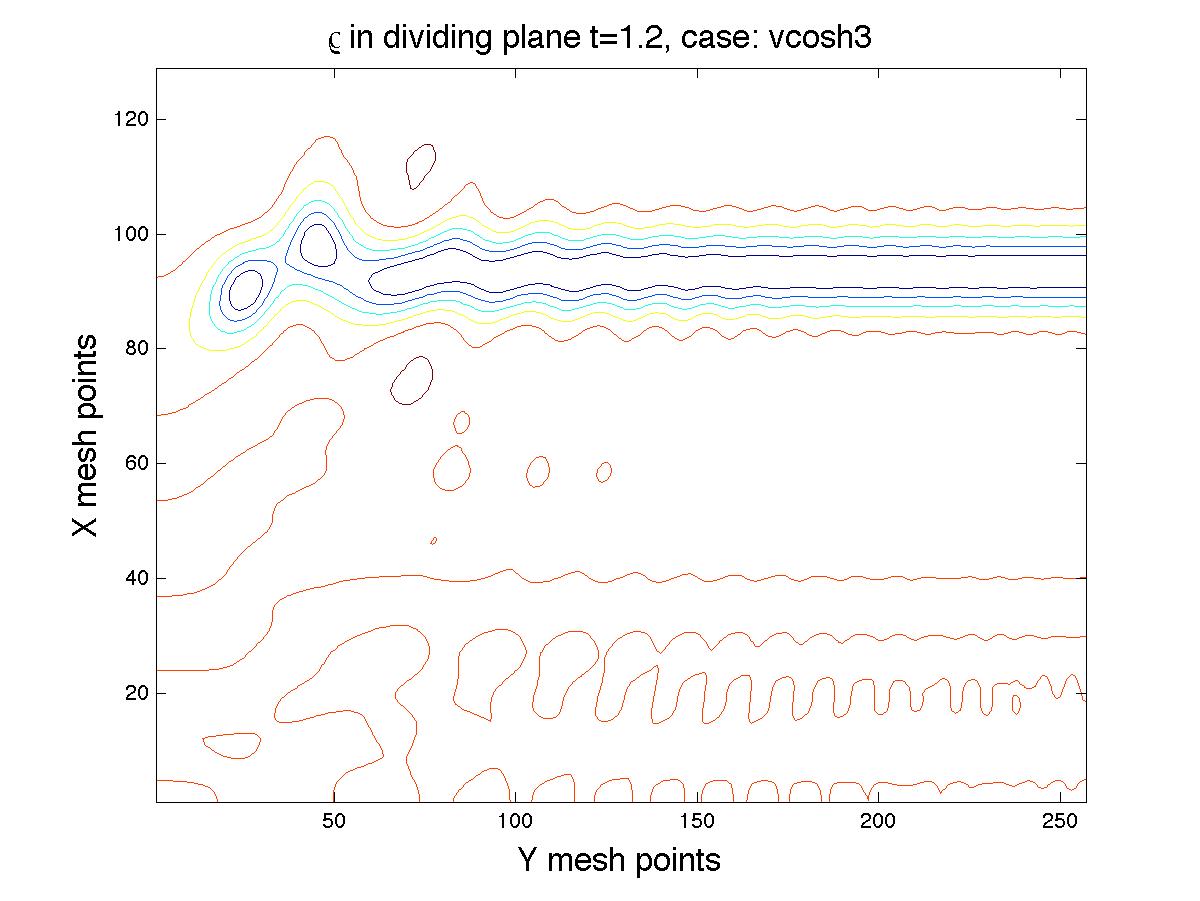}

\includegraphics[scale=.1]{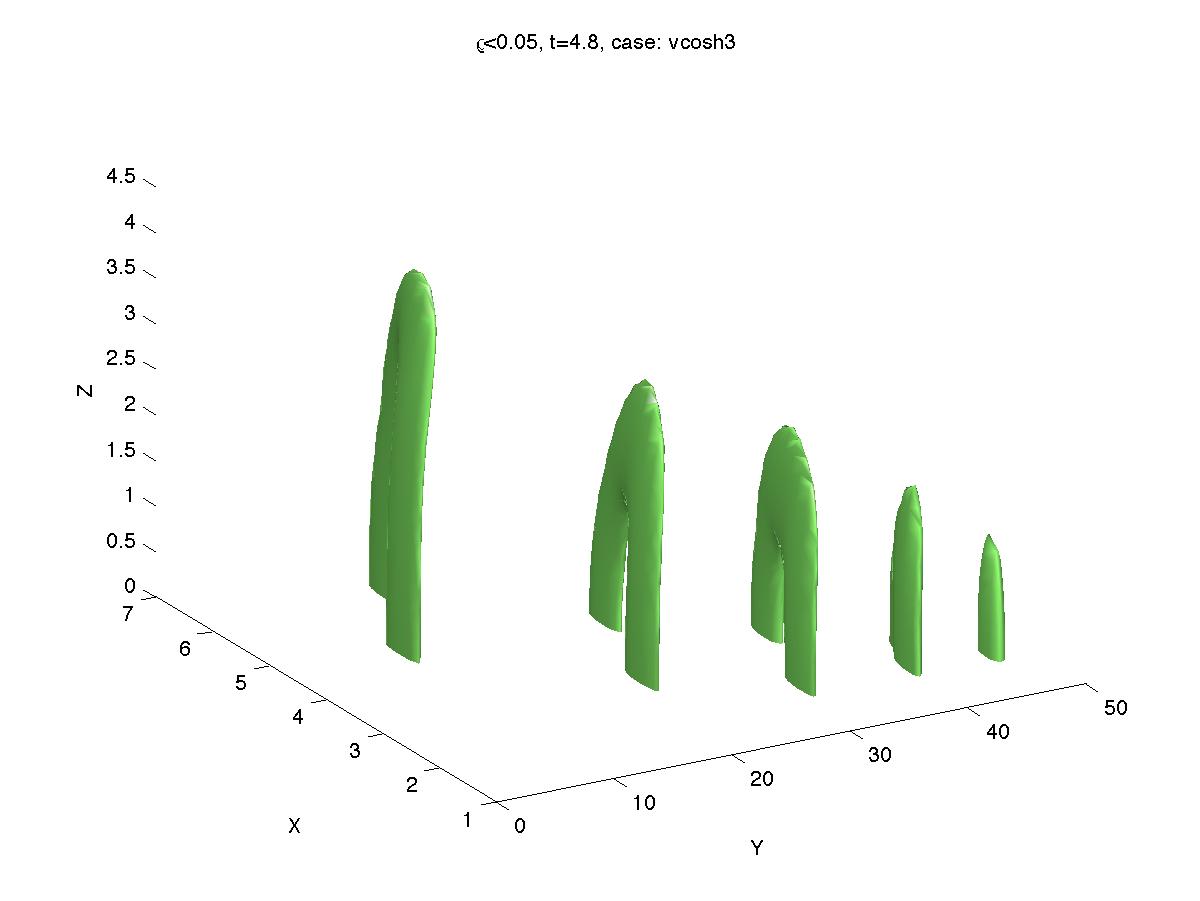}
\includegraphics[scale=.1]{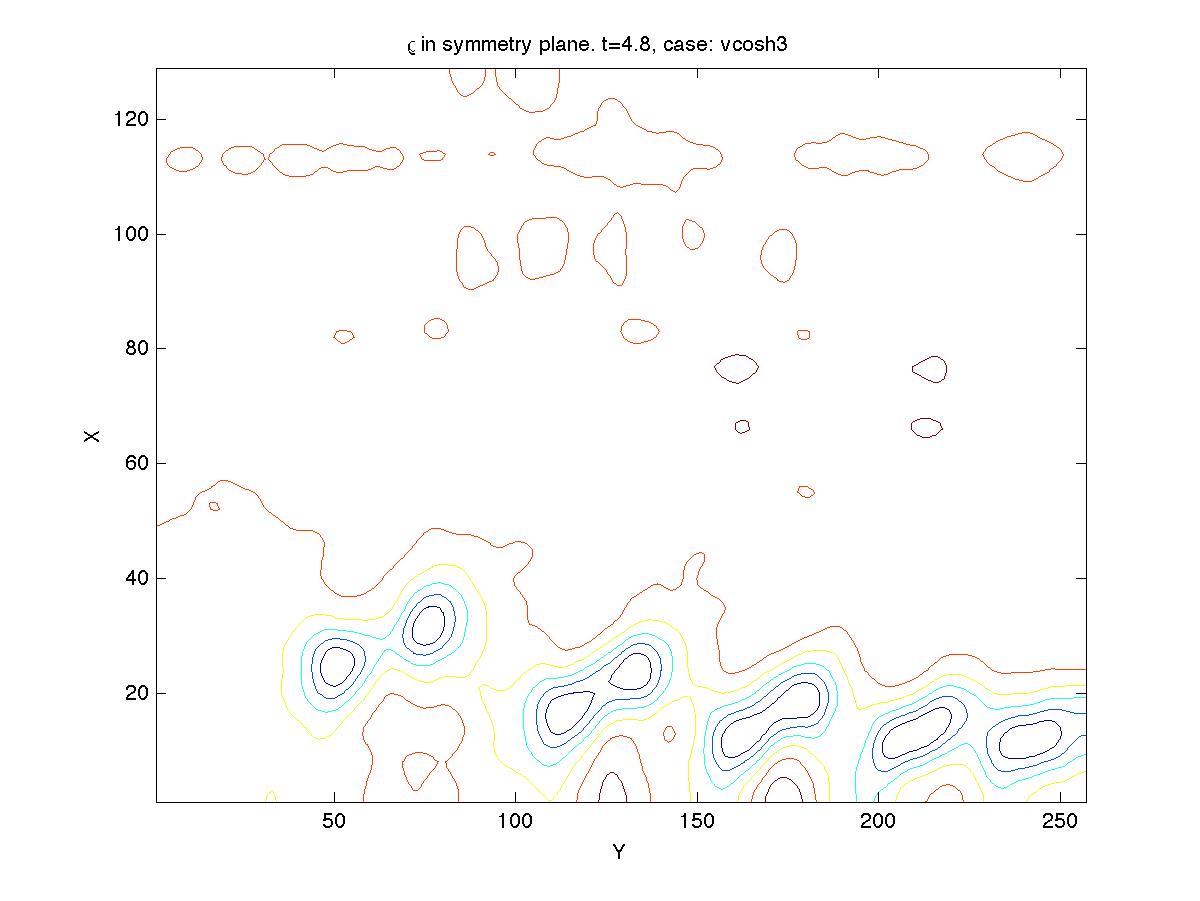}
}
\caption{Isosurfaces ($\rho<0.05$) and density contours in the $x-y$ 
dividing plane between the two original vortices. Times are
$t=12$ and 48.  Spectra at $t=12$, 26 and 66 are shown below.
}\EFG

\subsection{Spectra}

Before spectral results are presented, some caveats are in order.
Spectra are usually taken from large statistical ensembles, which
allows them to smooth out transients.  For time-developing calculations 
such as this calculation, smooth spectra are not expected.  The most
we can expect is to see a progression with time of energy to smaller or larger
wavenumbers. Therefore, the spectral slopes can be only approximate.  
Four spectral plots at successive (but not equally spaced) times are shown.  

What might be significant is that the kinetic energy moves to higher
wavenumbers in $k_x$ and $k_z$, the directions perpendicular to the
direction of the original vortex lines, and perpendicular to the
direction of propagation of the final vortex rings.  This would be
consistent with the cascade developing through the formation of
smaller and smaller vortex rings.

The interaction energy spectra are equivalent to spectra of the
density fluctuations, which have been measured in a superfluid \cite{Rocheetal07}.
The interaction energy spectra in these simulations are steeper
at all times than -5/3, suggesting that $E_I$  is being transferred to small
wavenumbers.

\BFG[!]\label{fig:earlyspectra}
\figcomment{
\includegraphics[scale=.18]{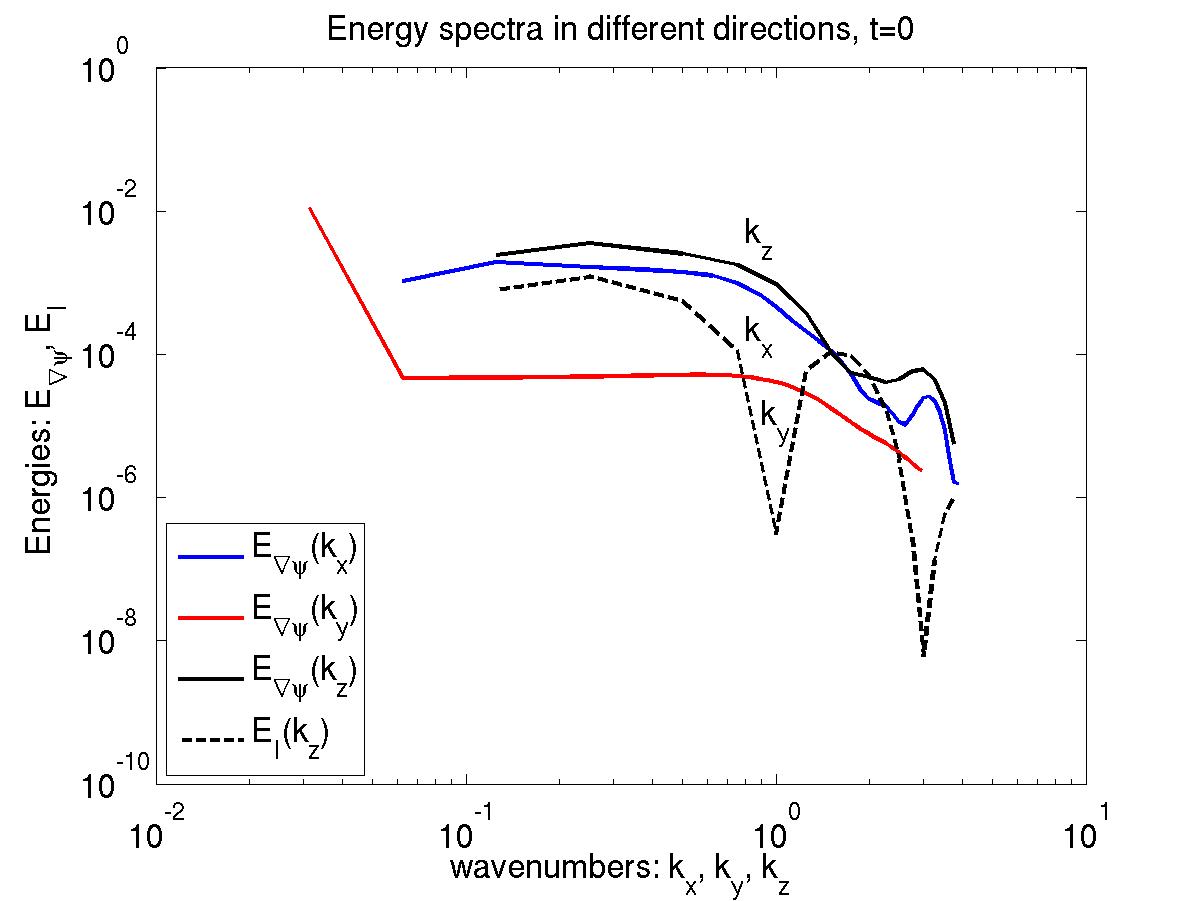}
\includegraphics[scale=.18]{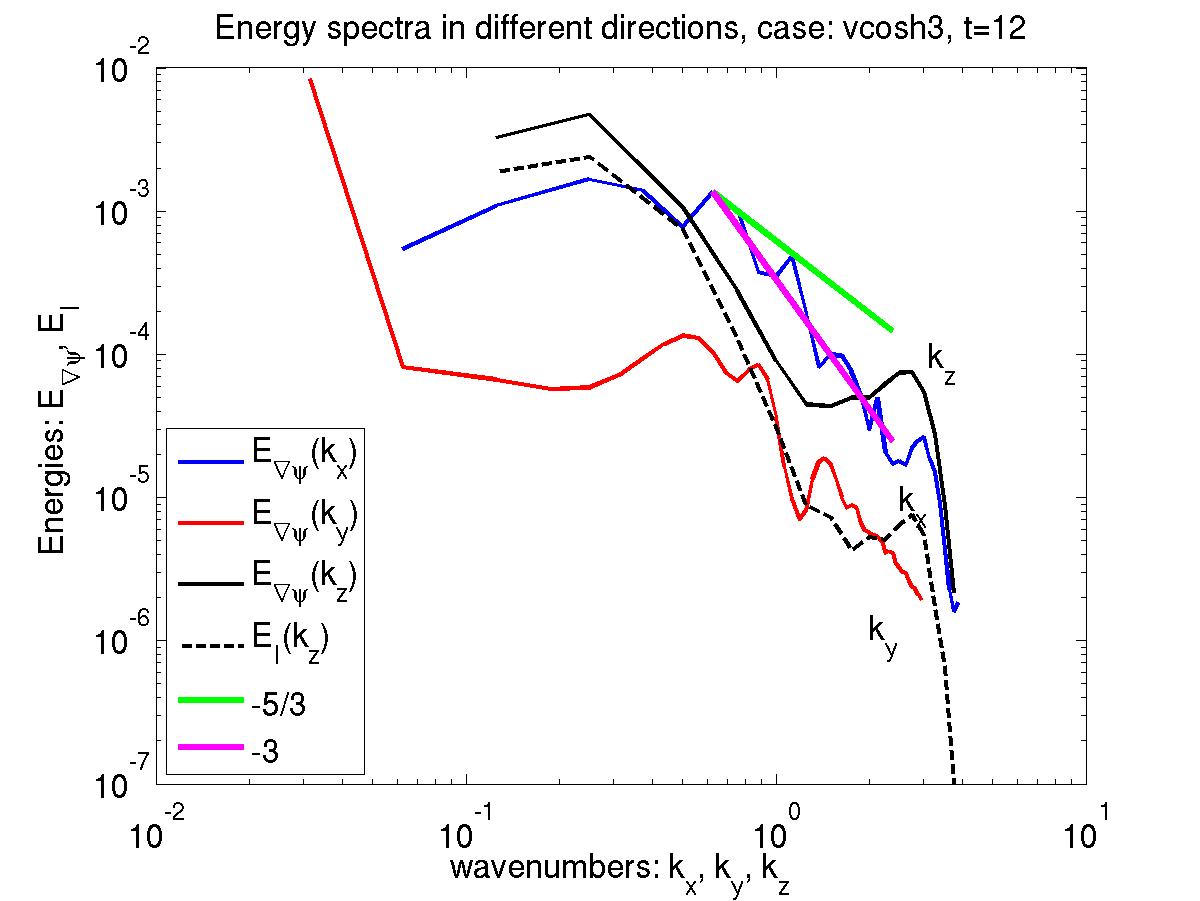}
}
\caption{Top frame: The initial one-dimensional
$K_{\nabla\psi}(k_j)=\sum k^2|\psi(k_j)|^2$ spectra in $(x,y,z)$
are shown, along with one interaction energy spectrum to demonstrate
that initially the two components of energy are not strongly correlated.
The initial spectra show resonances associated with the spacing, diameter
and the perturbation chosen.  This is especially true for $k_z$, where there 
is a bump at high wavenumbers that probably
represents strong gradients within the vortex core which is oriented
perpendicular to $x$ and $z$.\\
Bottom frame: By $t=12$, just after the first reconnection, the $K_{\nabla\psi}$ and $E_I$
components follow roughly the same scaling in all three directions and
to a rough approximation obey $k^{-3}$.  A strong
bump at high wavenumbers still persists from the initial condition for $k_z$.
While this looks like a bottleneck, it is only an artifact of the initial condition.}
\EFG

\BFG[h]\label{fig:midspectra}
\figcomment{
\includegraphics[scale=.18]{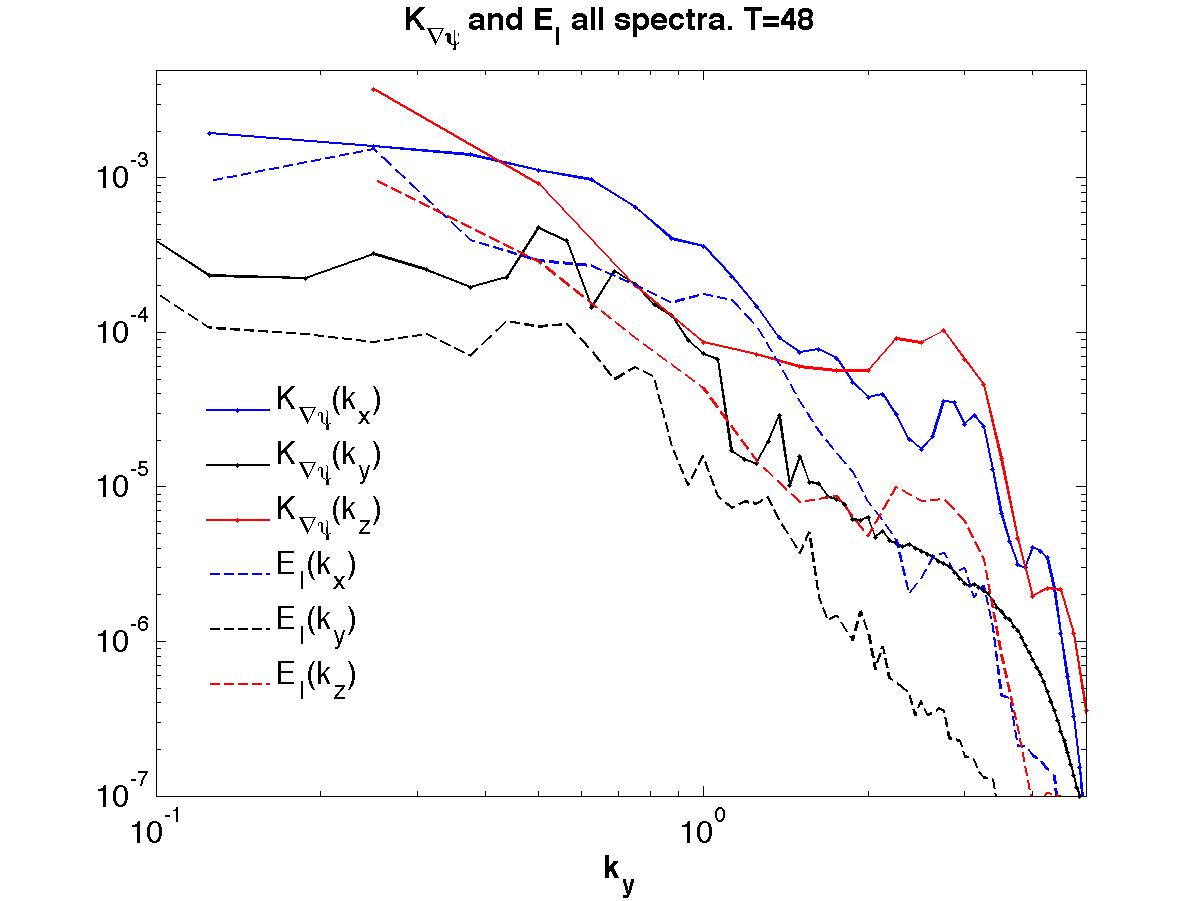}
\includegraphics[scale=.18]{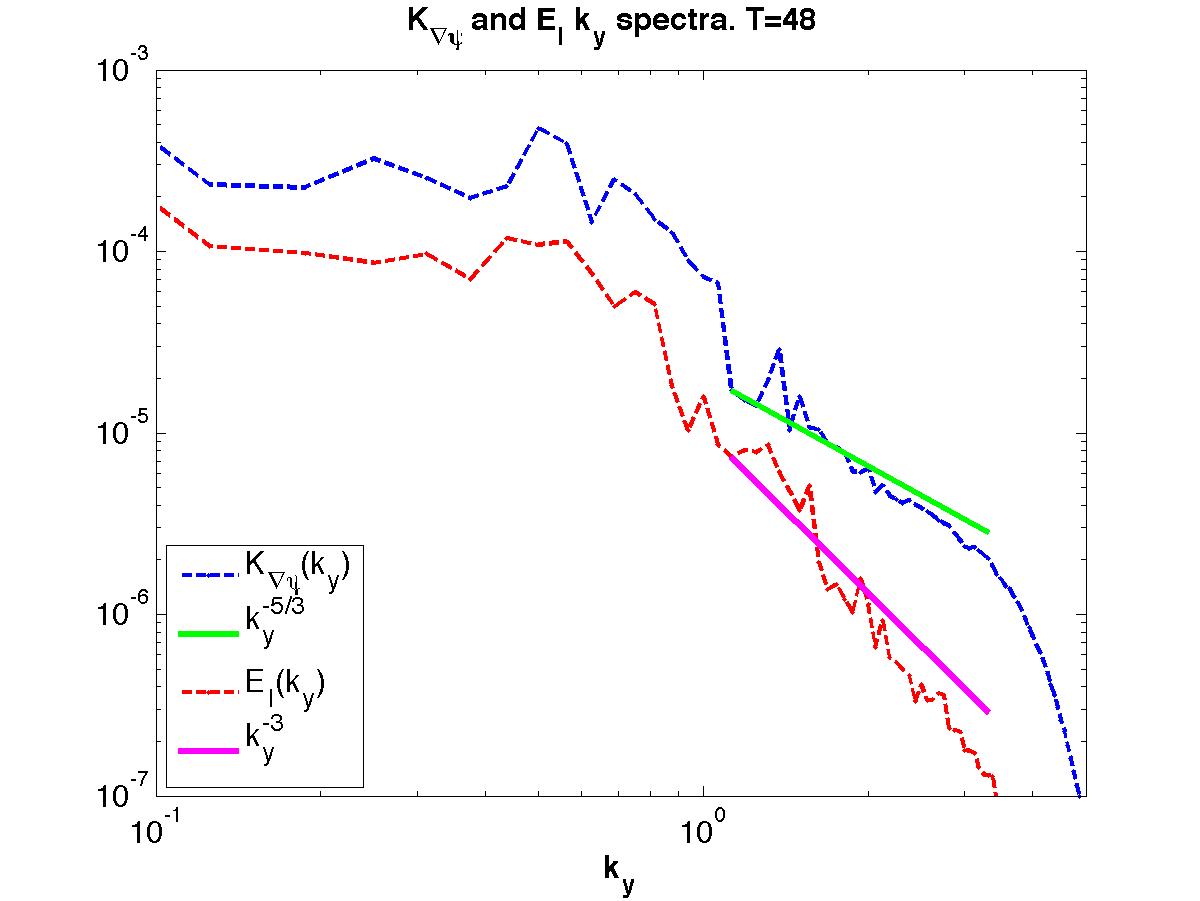}
}
\caption{Top frame: All spectra at $t=48$ for reference.  Bottom frame:
By $t=48$, for $K_{\nabla\psi}$ there is a long region in $k_y$ 
where the $K_{\nabla\psi}$ spectrum are the order of $k^{-5/3}$.
The interaction energy spectrum $E_I$ is still dominated by a $k^{-3}$ slope.
Spectra in the other directions have similar trends but are less distinct.}
\EFG

\section{Energy decay}

A final observational property of quantum turbulence that this 
simulation can address is how dynamics analogous to viscous dissipation
in classical turbulence could appear in an inviscid Hamiltonian system.
What is observed experimentally is the decay of the vortex line length.  
Or more exactly the density of deficits in the vortex cores, which scatter 
the second sound waves or ion beams used as probes
Assuming the lines are very thin, then the density of these
deficits is proportional to the length of the vortex lines.

\subsection{Assumed relation between line length and kinetic energy \label{sec:lineKE}}

This density of lines is usually interpreted as a sign of the decay of
kinetic energy by the following argument:
\ITM\item Vortex line length in superfluid turbulence is observed to decay as 
$\dint ds = t^{-3/2}$,.
\item If each quantum vortex line segment is viewed as an element of 
vorticity $\omega$, this implies that $\dint dV \omega^2 \sim t^{-3}$.
\item In classical turbulence,  $\dint dV \omega^2$ is known as the enstrophy,
which (multiplied by viscosity) is proportional to the dissipation rate
$\epsilon$ of the kinetic energy $E$.  
\item[] So if $dE/dt=\epsilon \sim t^{-3}$, then $E\sim t^{-2}$.
\item In classical, homogeneous isotropic turbulence in a periodic box the 
following is observed:
\ITM\item Energy: $E(t)\sim t^{-2} \quad\Rightarrow$
\item $ \overline{\omega^2}\sim t^{-3} $
\item \cite{Kerrthesis81}; but originally due to S. Patterson.
\ITN
\item Note that this decay law is never seen experimentally, as in all classical
experiments there are boundary layers.  It has only been seen in simulations with
either periodic or Neumann boundary conditions.  As explained above, the Neumann
boundary conditions used in the present calculations are a reasonable first-order
approximation to the true experimental superfluid boundary conditions.  So these
ideal classical simulations are relevant.
\item This is the basis for the claim that the observations of the
decay of vortex line density \citep{Smithetal93,WalmsleyGolov08} in
superfluids are related to the decay of kinetic energy in classical turbulence.
\ITN

When originally found in counter-flow  at higher temperatures \cite{Smithetal93},
the proposed explanation was transfer of energy to the normal fluid component
through mutual friction.  This would have been consistent with the
explanation for relaxation in rotating bucket experiments going back
to the 1950s.  However, it has been  pointed out that for this mechanism to
work in a counter-flow experiments, there needs to be sufficient time for
the superfluid and normal fluid vortices to equilibrate (Schwarz, private
communication, about 1995).

The more recent, lower temperature experiments 
\citep{Bradleyetal05a,WalmsleyGolov08} can completely rule out the
normal-fluid coupling explanation.  So how else could energy
be removed from the superfluid?  

\subsection{How could energy be depleted?}

All explanations center upon
the boundaries.  Unlike the ideal boundaries of these simulations, 
the physical boundaries in a superfluid cell are not ideal and
will remove energy, although not through viscous boundary layers as in
a classical fluid.  Therefore, the question of the origin of superfluid decay reduces
to finding a sufficiently efficient energy transport from the interior
to the boundaries to explain the observed decay of vortex line length.

Three mechanisms have been proposed.  
\ITM\item[1)] 
Quantum vortex lines could reconnect, form vortex rings, which then
propagate out of the superfluid tangle \citep{Feynman55}. 
Some role for this mechanism is found here.

\item[2)] Linear waves, or phonons could be generated and propagate
to the boundaries if their production is strong enough. The importance
of such a mechanism is demonstrated here, but not production of phonons
by vortex motion.  What is observed here would be consistent with
the production of rarefactions and phonons in calculations of the
collision of two vortex rings \citep{Leadbeateretal01} and other 
configurations \citep{Berloff04}.

\item[3)] Recently, there have been several proposals based upon the propagation
of waves on the vortices, which Fig. \ref{fig:waves} shows can develop on
Gross-Piteavskii vortices.  These wave theories could provide the decay mechanism if 
they transport energy sufficiently rapidly to their ends, presumably on the boundaries, 
or if there is a wave cascade to small scales along the filaments, at 
which point the smallest waves could generate phonons, which would then
move the energy to the boundaries.

For the Kelvin wave explanation to work, the cascade rate must be
sufficiently efficient.  Traditional wave cascade models have not
been able to achieve this efficiency, as pointed out by \cite{Lvovetal07}.
Alternatives have been proposed \cite{KozikSvis04,Laurieetal10}
based upon adding additional physics
and assumptions that would be difficult to verify experimentally.
\ITN

\subsection{Kinetic energy depletion in the calculation}

\begin{figure}
\includegraphics[scale=.4]{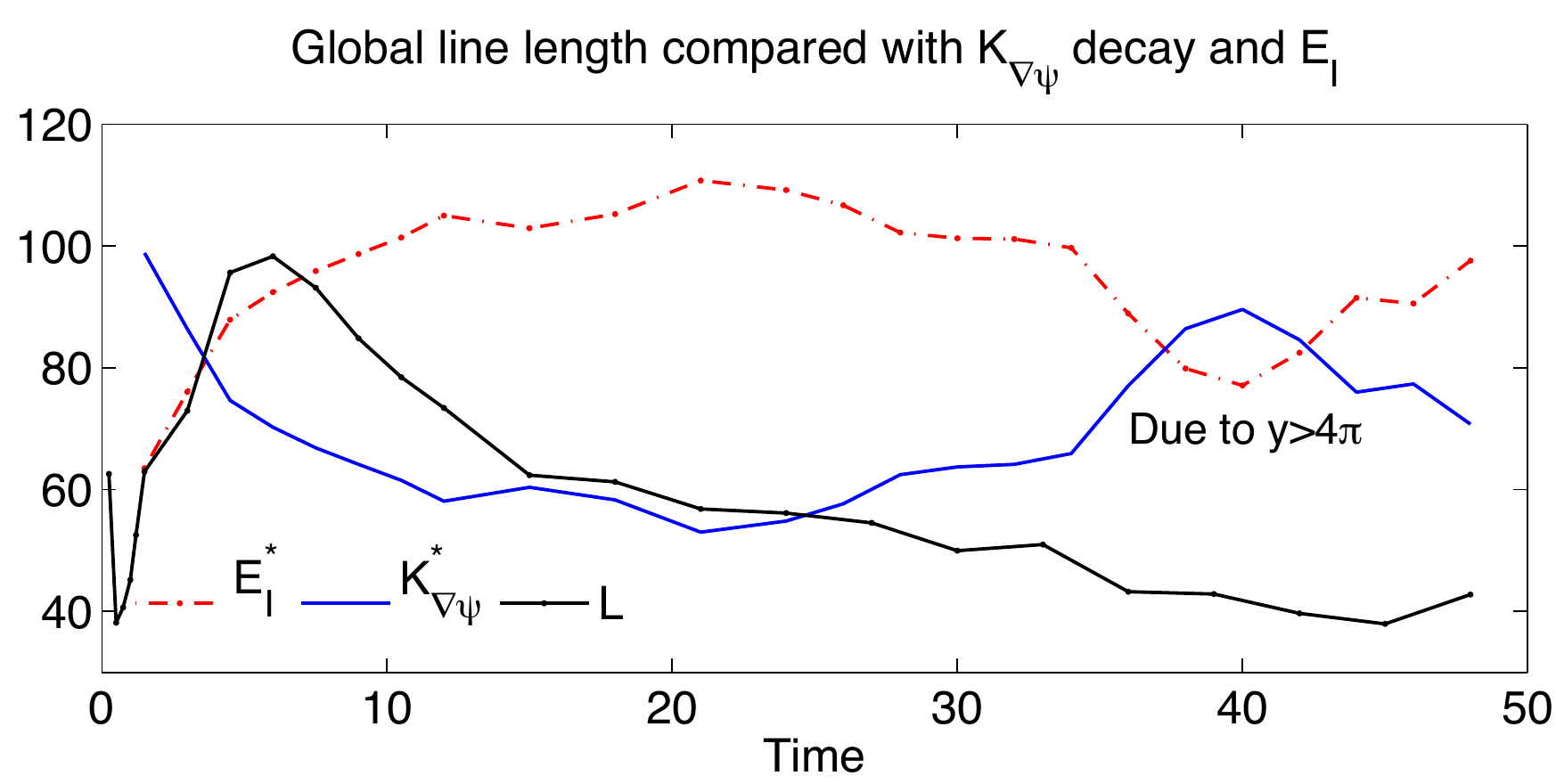}\\
\includegraphics[scale=.4]{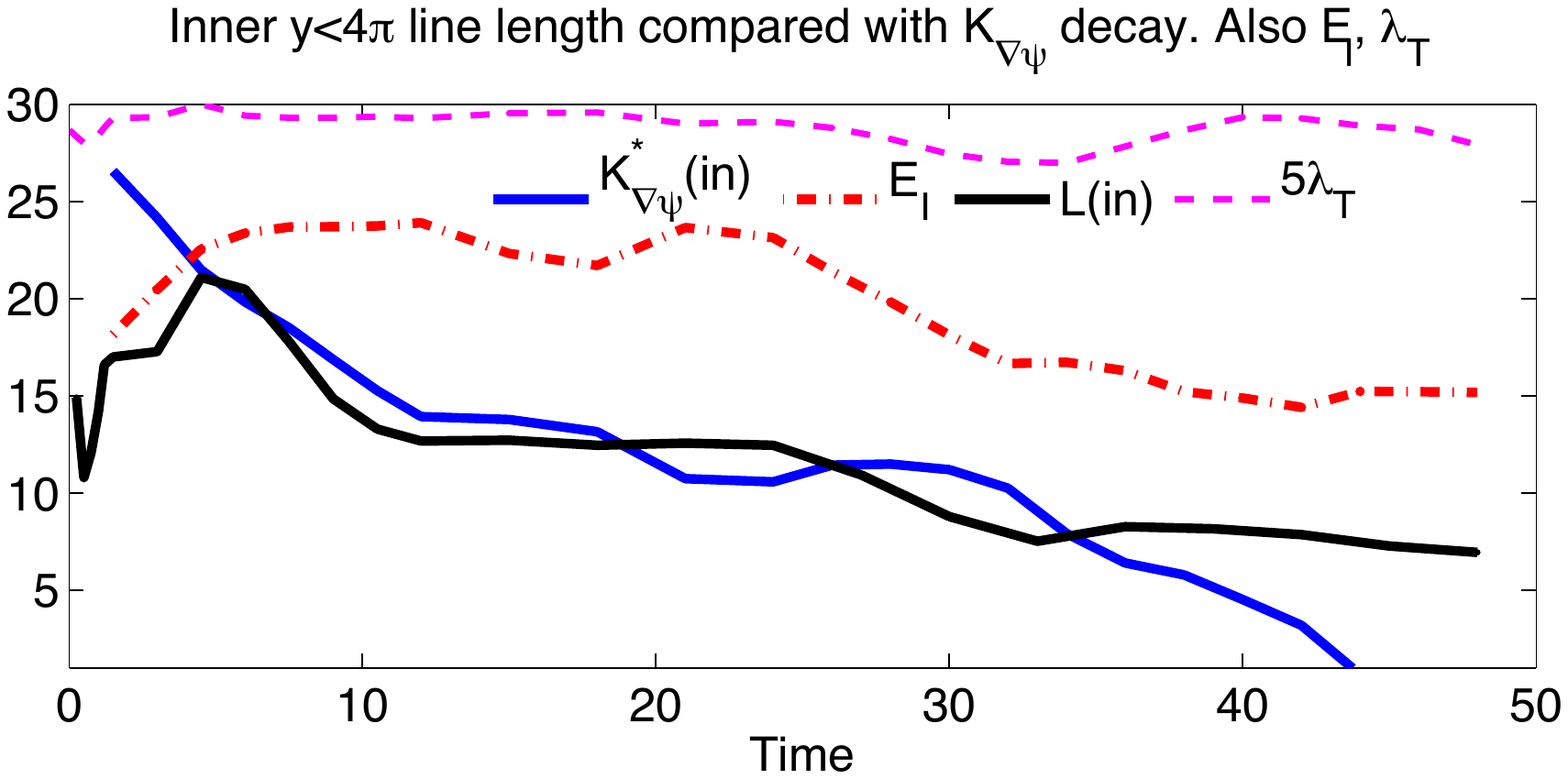}\\
\includegraphics[scale=.4]{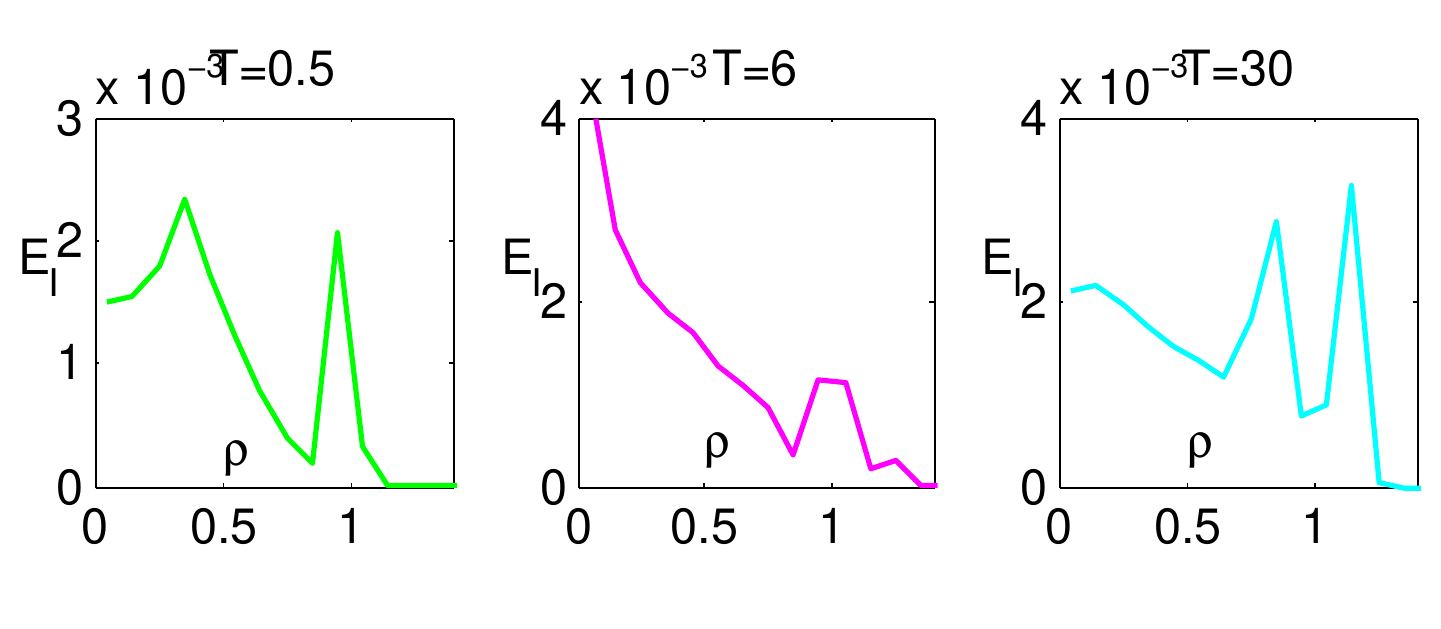}
\caption{Estimates of the line length compared to changes in
the interaction and kinetic energies.  Followed by distributions
of the interaction energy with respect to the density at three
times noted in the upper half at $T=0.5$, $T=6$ and $T=30$.
Analysis was done over the full domain (upper frame) as well as within
quadrants defined in the $y$-direction, with only the first $y$-quadrant 
plotted, which is where the interactions began.  
There is strong vortex line growth for $T=0.5-6$, followed by a period 
for $T=6-25$ of strong decrease in $L$.
Distributions of the $E_I$ with respect to density are shown
at $T=0.5$, $T=6$ and $T=30$ 
changes in the distribution of $E_I$ during these stages.  For
$T>30$ the global kinetic energy $K_{\nabla\psi}$ grows again.
This is associated with the accumulation of energy for large $y$
and the rings colliding with their images at the opposite wall.
In a physical cell, this energy would be absorbed by the outer wall.
To show that kinetic energy and line length continue to decrease in the
original interaction region, the middle frame shows 
$K_{\nabla\psi}$ and $L$ only in the first quadrant.  The final dashed
curve is a quantum equivalent of the Taylor microscale $\lambda_T$.}
\label{fig:linel}
\end{figure}

How can direct simulations of the Gross-Piteavskii equations help? 
The question is how to make the phenomenological relationship of the line 
length decay to the kinetic energy decay more robust. With a calculation,
all the components of the energy can be calculated directly, which we want
to related to a consistent, robust algorithm of determining the vortex line length.

Two methods have been proposed for determining the line length in calculations. Either
by counting boxes with a property associated with the vortex cores \citep{Leadbeateretal03}.  
Or by making the assumption that the spectral integral 
$\epsilon=\sum k^2 K_{\nabla\psi}(k)=\sum k^4|\psi(k)|^2$ 
is proportional to the line length squared \citep{Noreetal97} by the arguments 
above under Sec. \ref{sec:lineKE}.

The analysis below shows that with the box-counting proposed here, the
development of the anti-parallel calculation might provide the first
numerical evidence of when both line length and
kinetic energy in the original interaction region decrease.
Elements of this can be found in earlier work, but either the box counting
analysis is not compared to the kinetic energy \cite{Leadbeateretal03,Berloff04} 
or the spectral line length analysis is unphysical.  That is $\epsilon$ 
grows while kinetic energy decreases \citep{Noreetal97}, or in the present calculations
$\epsilon$ does not decay.

The box-counting criterion chosen here is to a pick low density threshold $\rho_c$ that
will estimate of the volume occupied by the vortex cores. By dividing this volume
by the area of the analytic profile \eqref{eq:initprofile} with $m=1$, and choosing different
density thresholds, consistency checks can be made. 

There are two primary stages to the time dependence of the line length.
After a brief adjustment from the initial condition, completed by $T=0.5$,
the line length grows, which would be consistent with vortex stretching.
The stretching stage lasts beyond the first reconnection, near $T=1.25$,
and on through the second reconnection near $T=4.5$ when the first and
largest vortex ring separates off, as illustrated by Fig. \ref{fig:waves}.

Thereafter the line length decays.  To understand the different stages,
below the main figure there are three subplots showing distributions 
of the interaction energy $E_I$ \eqref{eq:GPenergies} with respect to density
at different times.  If vortex stretching is 
occurring, one might expect there to be a strong growth in regions with 
very small densities.  If the energy is in waves, then we might expect
the distribution of $E_I$ to be concentrated around $\rho=1$.  Around,
not at, because for $\rho=1$, $E_I\equiv 0$.  Distributions of the
density are not useful since their maximum is peaked near $\rho=1$
for all times.  
The distribution at $T=0.5$ is meant to demonstrate that initially,
$E_I$ is maximum near $\rho=1$.  

During the period of vortex stretching there is a dramatic growth
in $E_I$, with most of the growth at small values of $\rho$, as demonstrated
by the distribution for $T=6$ when the line length $L$ is maximum.
Because $E_I(\bx)=(1-\rho)^2$, small $\rho$ at a given $\bx$ automatically
implies a small value for $E_I$ at $\bx$, so what the distribution at $T=6$ is telling
us is that there is a large growth in the number of points with
$\rho\approx 0$.  Note that the strong increase in $E_I$ during each of these stages is
compensated for by a strong decrease in the kinetic energy 

Immediately after $T=6$, $L$ begins to decrease dramatically while
kinetic energy $K_{\nabla\psi}$ continues to decay, which is
compensated for by a continuing increase in the interaction  energy 
$E_I$.  At the end of this stage, there is a growth in large values
of $E_I$ on either side of $\rho=1$, shown by the distribution
at $T=30$. This would be consistent the development of waves and
the visualizations of waves being emitted from two colliding vortices 
\citep{Leadbeateretal01,Berloff04}.  
The double humped distribution relaxes gradually for $T>30$.

A decrease in the global kinetic energy does not persist.  Eventually
interaction energy converts back into kinetic energy.  This could be associated
with the structures (rings) formed by the multiple reconnections
reaching the outer $y$-boundary, or due to the existence of a wave-dominated state. 
Similar oscillations where observed
in GP calculations with a symmetric Taylor-Green type of initial condition
\citep{Noreetal97}.  This would not happen in a real 
experimental device because the boundary conditions are not ideal.

To mimic what decay in a real flow might look like, especially
experiments that generate tangles far from boundaries \citep{WalmsleyGolov08},
the final two curves in Fig. \ref{fig:linel} show the growth, then decay, of
the line length in the first $y$-quadrant ($0\leq y \leq4\pi$), and
a rescaled kinetic energy decay curve. When viewed in this way, the
local region exhibits kinetic energy depletion due to transport out of that
region in physical space after the global kinetic energy begins to reappear.

\section{Analysis and calculations in progress}

For this calculation to be accepted as a paradigm for the dynamics of kinetic energy
decay in quantum turbulence, it must be demonstrated that the events
seen here are not unique to this particular initial condition.  The
four ring case noted below would support this claim.  What the
anti-parallel case provides that more general cases cannot provide
is a tool for addressing deeper questions about the cascade dynamics, and
in particular the relationship with the creation and destruction
of physical space structures.  What should be addressed?

\subsection*{Nonlinear spectra: Formulation and cascade trends}

An unfinished piece of analysis is the Fourier spectra of the energy transfer.
The equations have been formulated, and one property is that the
kinetic energy and interaction energy must mediate the cascade of the other.
There cannot be a cascade of just kinetic energy or interaction energy. 
They must be coupled.  Which directions do the two cascades go?  The
evidence so far suggests that there is a cascade of kinetic energy
to small scales, while the conversion to interaction energy is through
a coupling with much smaller wavenumbers (i.e. large length scales).

\subsection*{More general initial conditions}

Upon presentation of these results at several meetings, one
consistent concern was whether the dynamics observed here are
an artifact of the particular strongly anti-parallel initial
condition, or might be more general.

One of the goals of these additional cases will be to test the
validity of assumptions made by recent vortex wave models for
the quantum turbulence cascade.  So far, all wave growth
and all reconnections seen are driven only by interactions
between vortices that were originally distinct.  There has
been no evidence for any signficant wave growth or appearance
of reconnections due to individual vortex dynamics.  Rather,
the generation of $E_I$ due to stretching and the phonons that
are generated in the process seem to suppress all auto-vortex
wave and reconnection processes.  In particular there is no
evidence for cascades of waves appearing on individual vortices
as one model with higher order Biot-Savart terms predicts
\cite{Laurieetal10}.  And reconnections are never sufficiently
sharp to generate the torsion needed to drive self-reconnections
on vortices as proposed by another model \cite{KozikSvis04}.

\subsection{Orthogonal}

\BFG[!]\label{fig:ortho}
\figcomment{
\includegraphics[scale=.1]{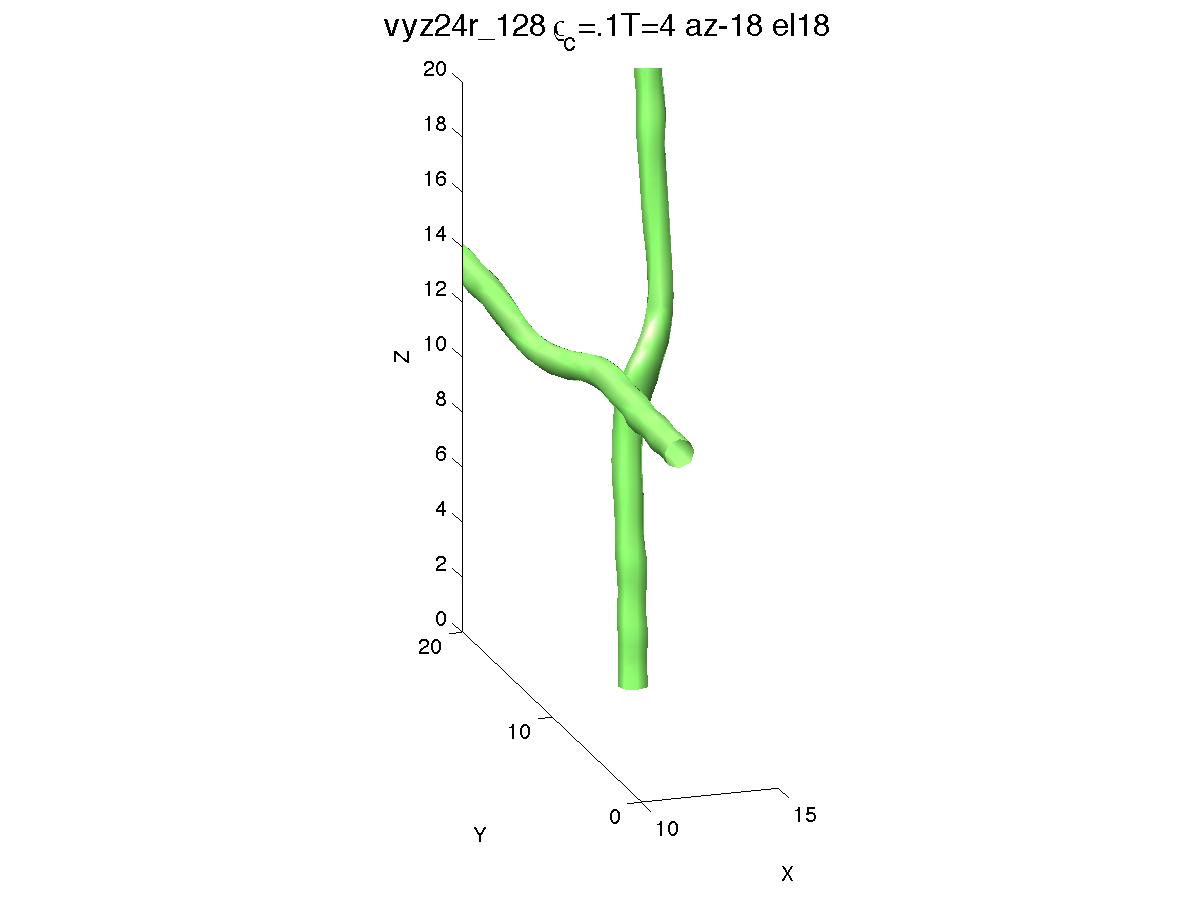}
\includegraphics[scale=.1]{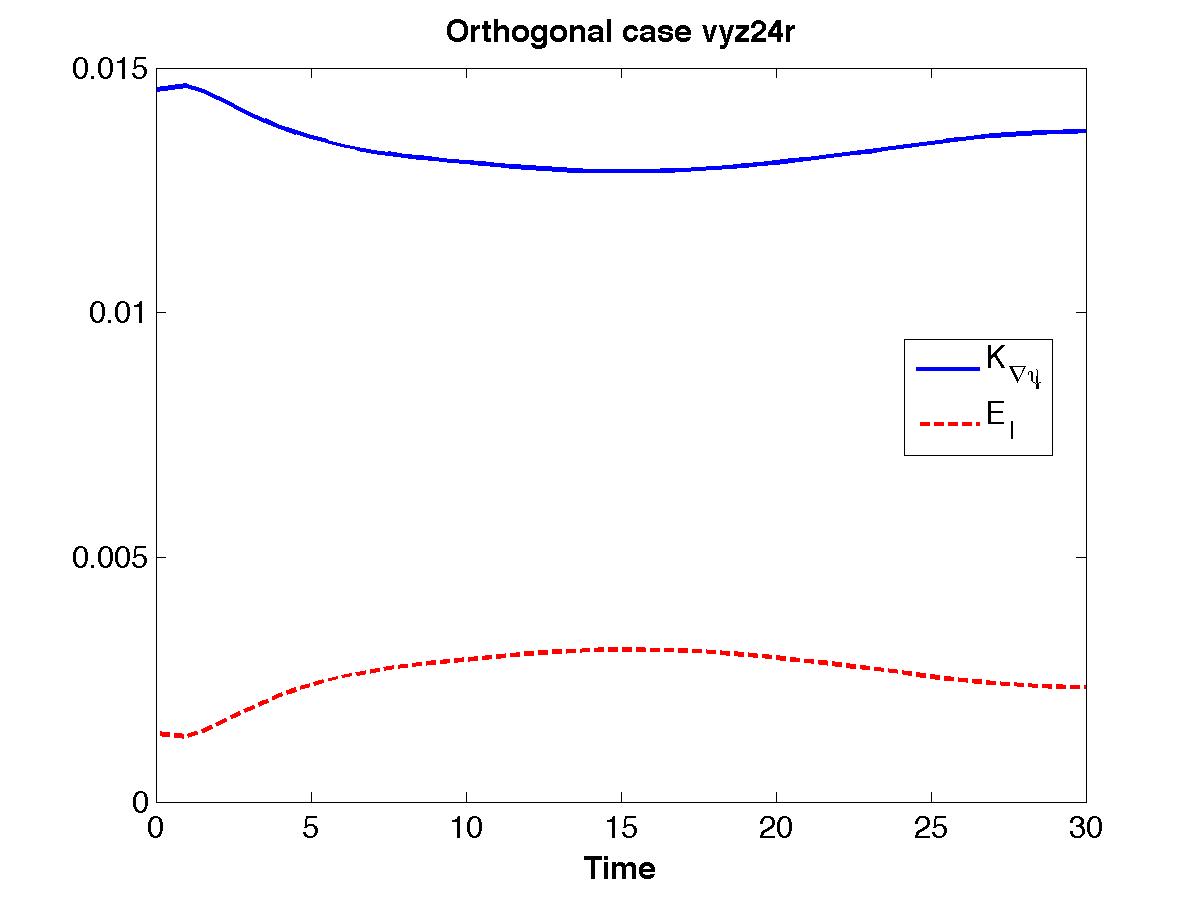}

\includegraphics[scale=.1]{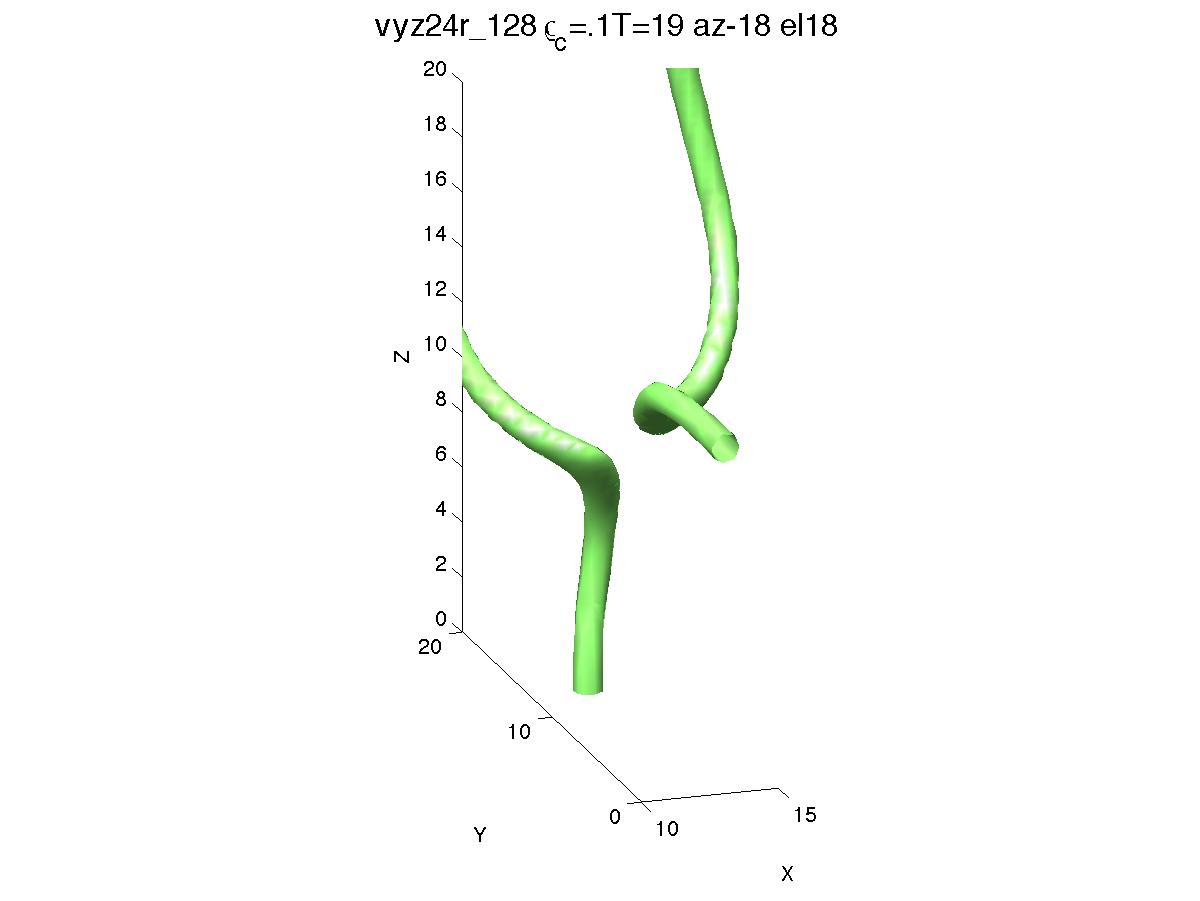}
\includegraphics[scale=.1]{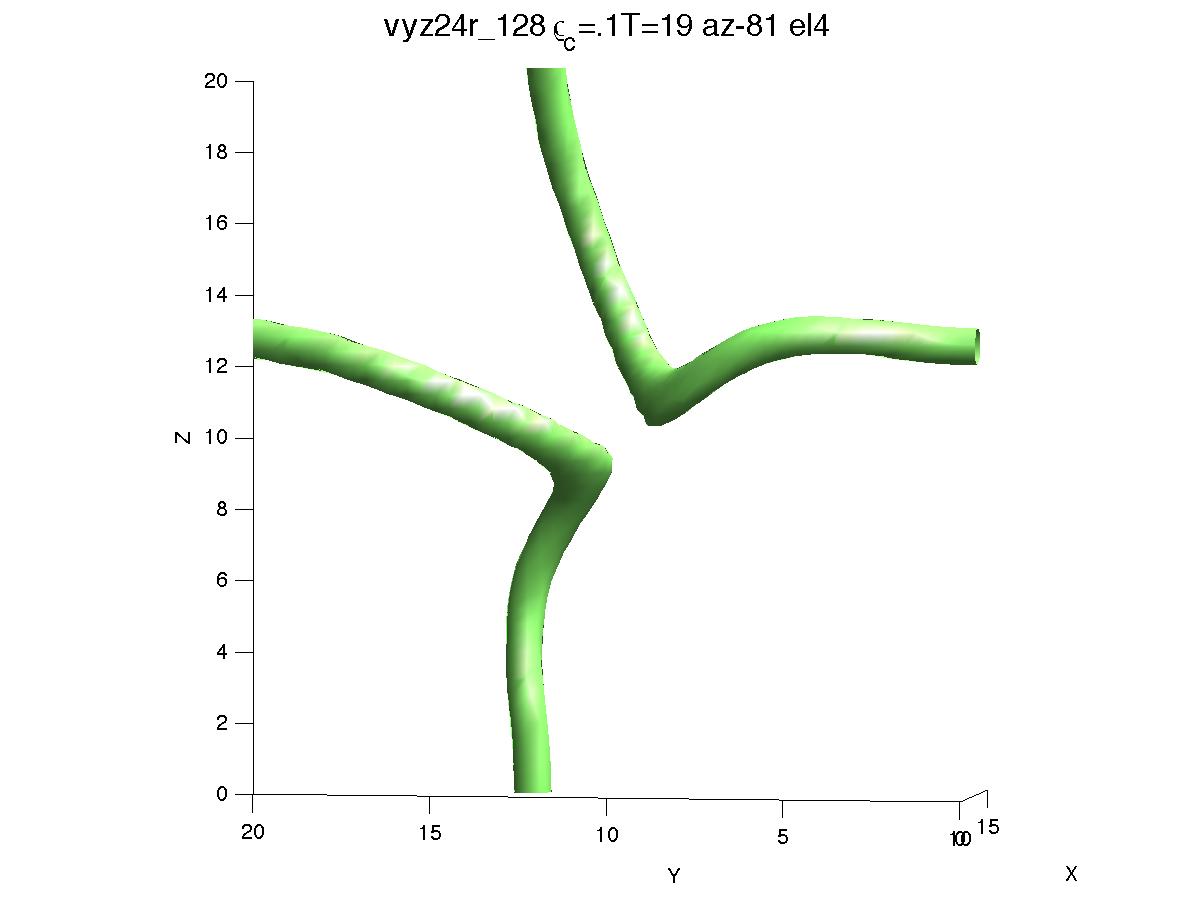}
}
\caption{Two orthogonal vortex liness, 
the time dependence of the kinetic and interaction energies, two views
just after reconnection.
}
\EFG

The case most frequently asked about is initially orthogonal vortices.
Test calculations were done and are essentially no different than
what was first presented in 1993 \cite{KoplikLevine93}.
The reconnected vortices are twisted, but not enough to generate 
waves or any further self-crossings and reconnections.
This is distinctly different from the LIA filament analysis of
\cite{Lipniacki00}, from which at least one additional one additional reconnection, 
would be predicted.  

This orthogonal case also resembles Case A from \cite{Bewleyetal_PNAS08} for
the experimentally visualized quantum vortices recently highlighted in
in Physics Today \citep{PhysTodayJul10}.

Based upon simulations of the reconnection of classical orthogonal vortices,
one would have expected there to be self-reconnections.
Further tests with the original vortices are intermediate angles need to be
explored to determine if this result is general.

\subsection{Two rings}

\BFG[!]\label{fig:2d_ring}
\figcomment{
\includegraphics[scale=.1]{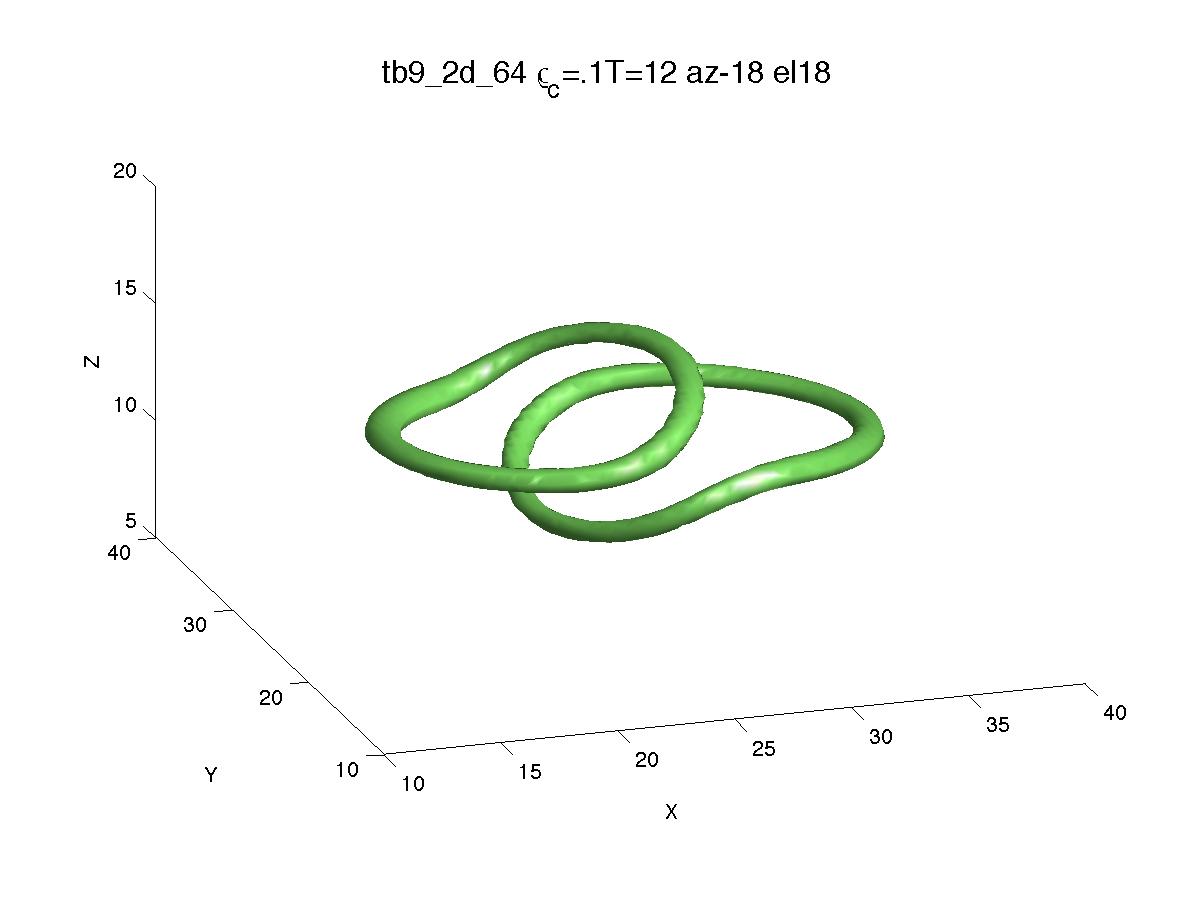}
\includegraphics[scale=.1]{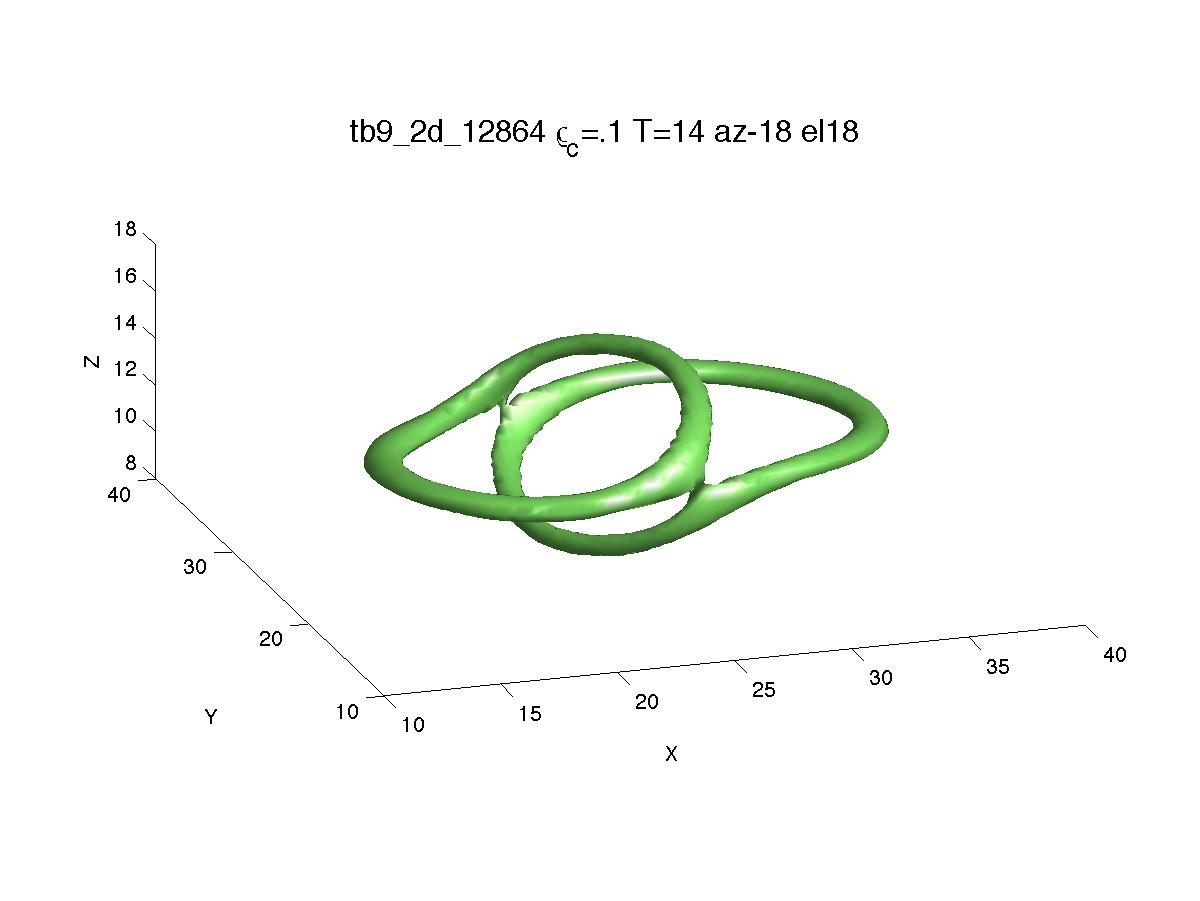}

\includegraphics[scale=.1]{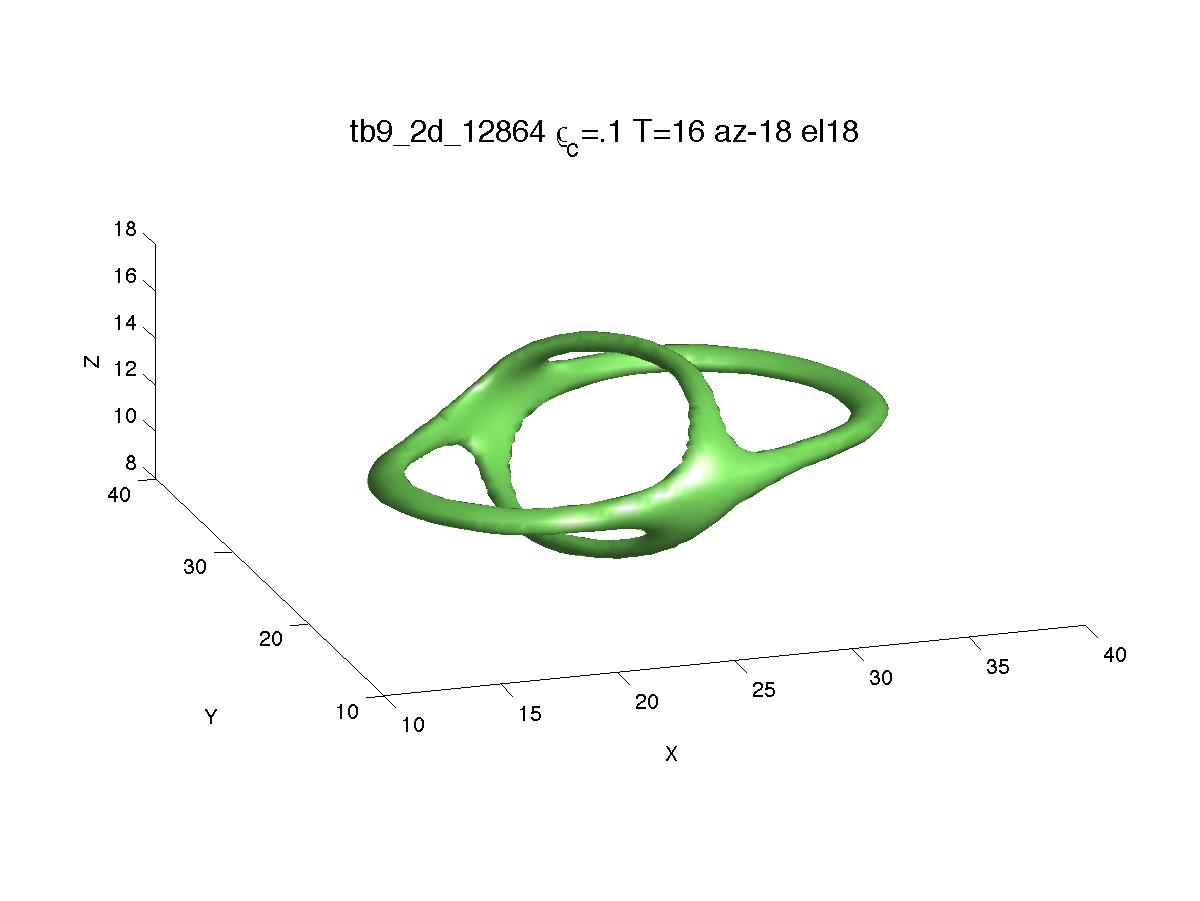}
\includegraphics[scale=.1]{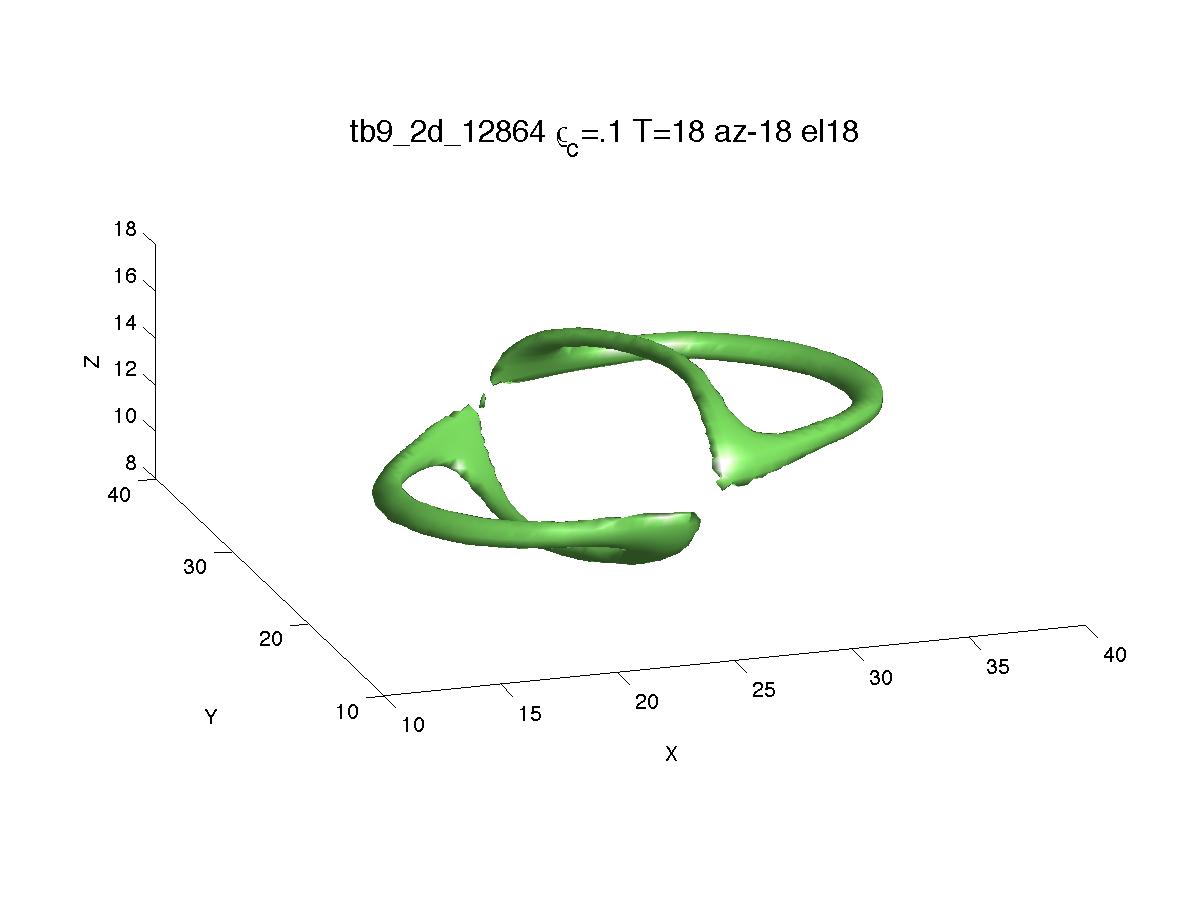}
}
\caption{Two vortex rings combine via reconnection, then separate.}
\EFG

In the search for a structure with collapse properties similar to 
the primary anti-parallel case here, a number of colliding ring
configurations were explored.  One batch of tests used approximately
anti-parallel pairs, all with their axes in the
$x-y$ plane, each with slightly different radii, different ellipticities
and displaced horizontally from one another.

All except one followed roughly the same scenario.  As they 
approached, first there would be two reconnections.  This would
then form a roughly figure-eight configuration of the vortex lines.
Then with further reconnections this would break down into two
outgoing vortex rings.  

There are similarities to the dynamics of two colliding rings with 
the same radii, and displaced horizontally, reported by 
\cite{Leadbeateretal01}. They do not get a figure-eight, but a more
closed intermediate structure.  However, they do get two rings
coming out of the intermediate structure.  The purpose of their 
paper was to demonstrate that phonons (or GP Kelvin waves) appear
from the reconnections and possible evaporation of the smaller 
outgoing rings.  This would be consistent with the anti-parallel case here.
Further analysis of higher resolution calculations
would be necessary to determine if this is happening for these test
initial conditions.

In one distinct case here, the combination of ellipticity and displacement of the
vortices was sufficient that they reconnected at opposite ends
of their ellipses, at the point where originally the vorticity was
parallel.  But first, one of the rings the vorticity twisted itself
around so that the reconnection was still between approximately
anti-parallel vortices.

\subsection{Four rings}

\BFG[!]\label{fig:4rings}
\figcomment{
\includegraphics[scale=.1]{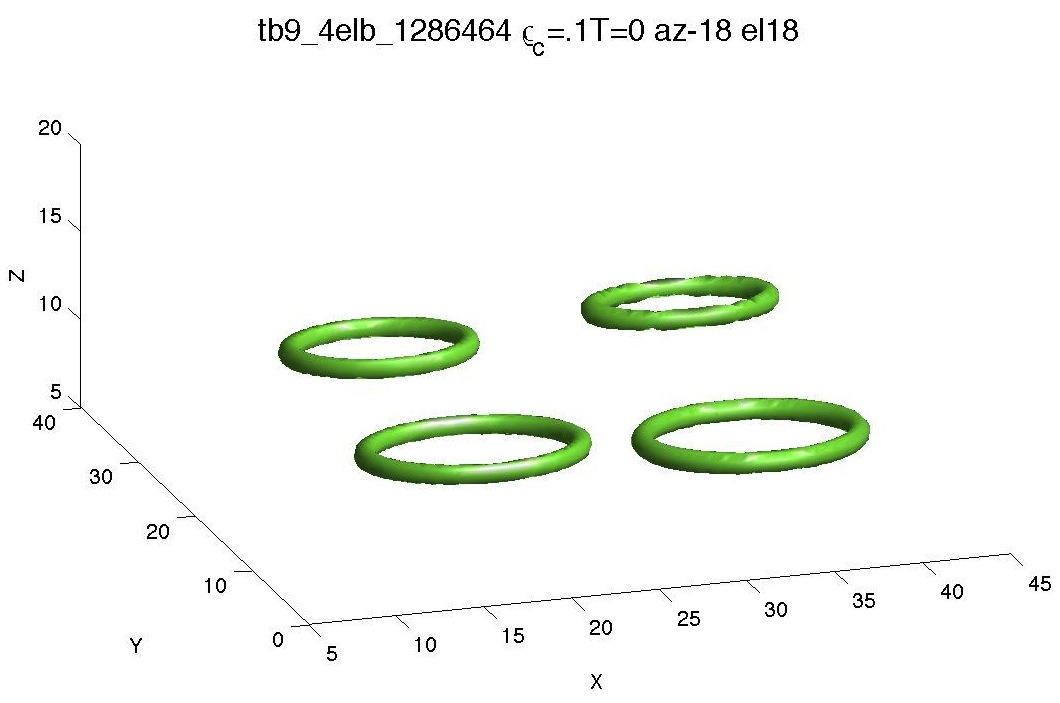}
\includegraphics[scale=.1]{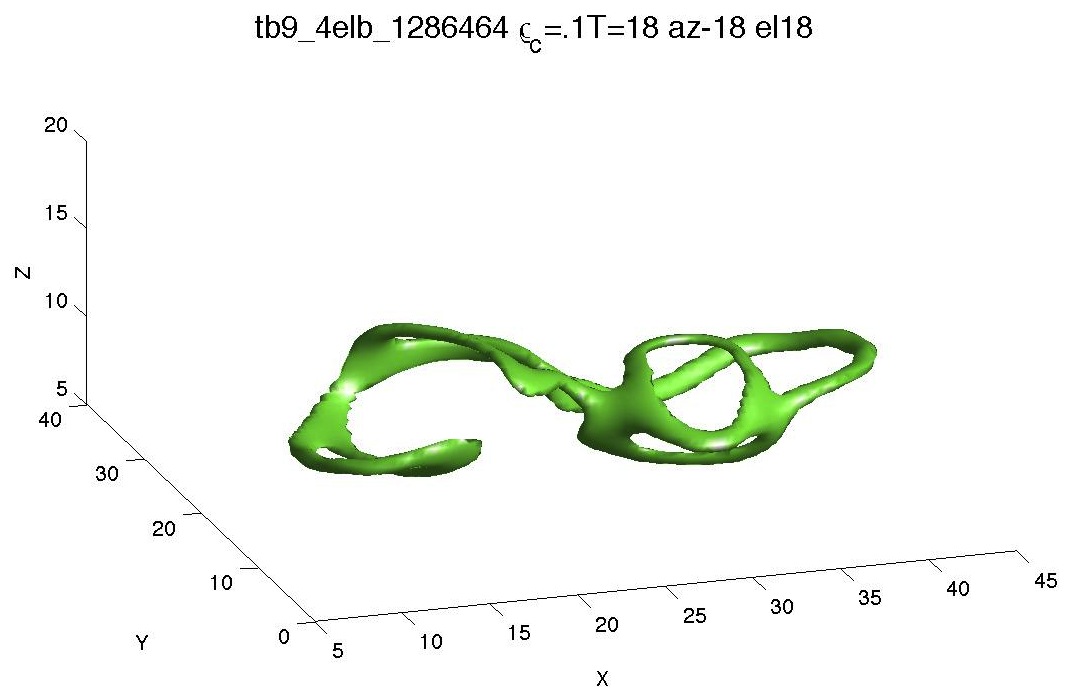}

\includegraphics[scale=.1]{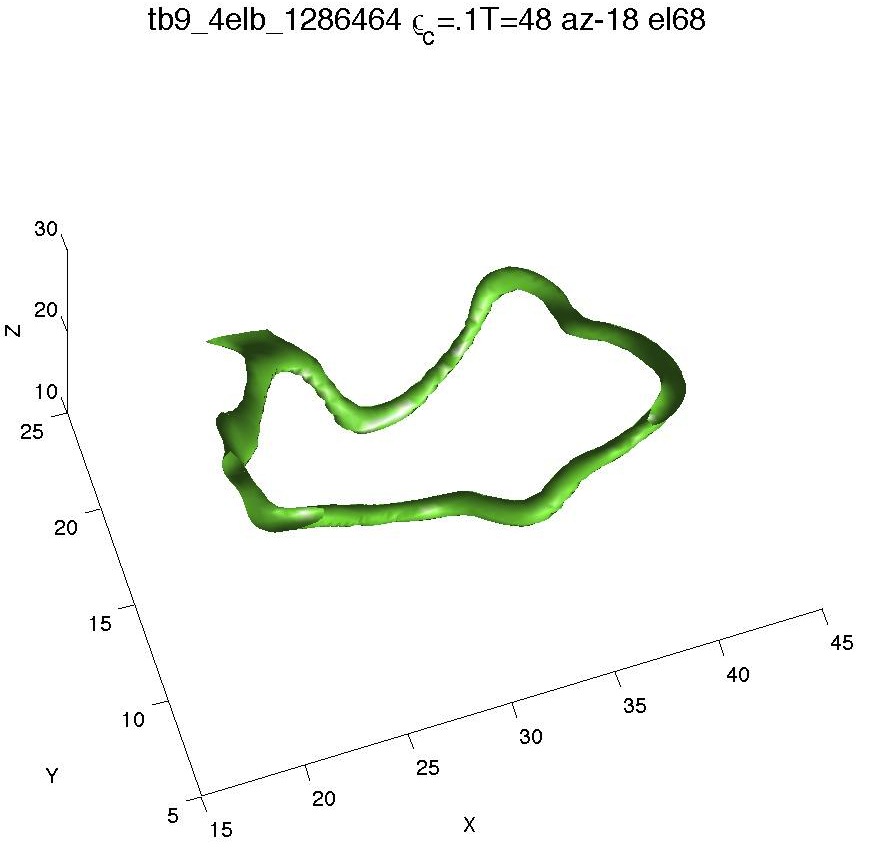}
\includegraphics[scale=.1]{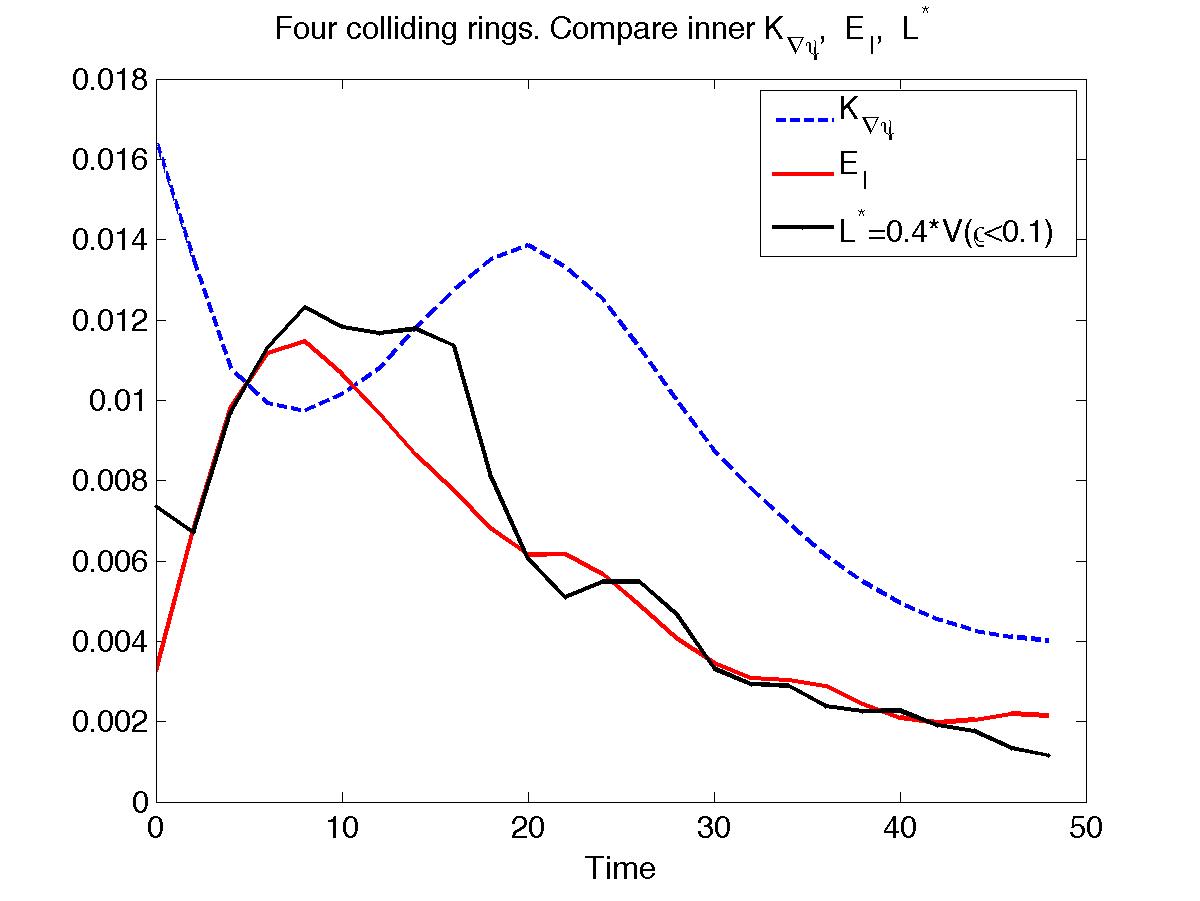}
}
\caption{Four colliding rings, the entangled state generated, the
final relaxed state, and the time dependence of the kinetic and interaction energies
plus a measure of line length in the inner region that contained the original
vortices.  In this case the measure of line length, the volume where $\rho<0.1$,
tracks the interaction energy $E_I$ more closely than the kinetic energy.
}
\EFG

The best generalization of the anti-parallel
calculations is to let four vortex rings collide, shown in Fig. \ref{fig:4rings}. 
These temporarily combine to produce of tangle of wavy vortices, which breaks up
into a variety of small rings that appear to evaporate (perhaps turning
into phonons) and a large vortex ring. The large ring probably removes 
most of the original Hamiltonian. 

Three figures show the initial condition, the state of maximum
entanglement, and the final state of a single large vortex ring. 
If the initial state is envisaged as two anti-parallel pairs,
to get the entangled state it was necessary that the separation
between the pairs be the order of their radii, but not overlapping.

The fourth figure shows the time development of the kinetic and
interaction energies.  Again there is a strong initial transfer of energy
from the kinetic to interaction components.  The entangle time
shown is roughly at the end of this process. After that, as
structures start interacting with the boundaries, the interaction
energy is converted back into kinetic energy.  Line length analysis 
is needed as well as local analysis around the region of the initial interactions 
to determine if some of the trends observed in the anti-parallel case
are repeated.

Before this is done, this case needs to be rerun, first with much higher resolution
to see the details of the entangled state, then with a combination of 
higher resolution and larger domain to verify that the final state
is not too strongly influenced by the boundary conditions.

There are similarities to the four ring case investigated by 
\citep{Leadbeateretal03}.  The four rings in that case are all
oriented perpendicular to a common axis and have more symmetry than
the case here.  Resolution is also less than half that used here.
Nonetheless, two primary wavy, vortices develop, plus smaller entities
that appear to break down into waves.

\section{Outlook}

For many years, simulations of first Navier-Stokes turbulence 
\citep{Kerr85} have been able to generate $k^{-5/3}$ spectra
and turbulent structures,
now getting 2 decades using up to $4096^3$ mesh points \citep{Kanedaetal03}.
More recently it has become possible to do the same with the 
Gross-Piteavskii equations \citep{Yepezetal09}.

However, there are no clear relations between the spectra and
structures in any of these cases.

What would be particularly useful would be a single case, no
matter how special, that could demonstrate a connection between
vortex structures and the origin of the energy cascade.

There have been special cases in the past showing a possible
connection between structures and spectra, for example the
Lundgren vortex \citep{Lundgren82}, but that case requires
the existence of an imposed large scale stretching.  More
interesting would be a case that generates a flow of energy
across scales that starts with individual structures.

The case presented here provides such an example, and with
time we should be able to demonstrate most of the underlying
steps in more general cases similar to the final
example given.  Even then, the value of the special
anti-parallel case is that it will be easier to identify the dynamics.

Some of the dynamics discussed here have been noticed before
as referenced above.  This includes initial development where the
length of the vortex lines grows, that is the existence of vortex stretching,
the generation of waves on vortices, reconnection between lines and rings, 
and the determination of spectra.

New relations between some of the known properties are noted here.
For example, waves following artificial reconnections in the Biot-Savart and local
induction approximations have been seen, but waves on individual quantum vortices 
following reconnection have not been seen before.  And from that, one can see
is how those waves can lead to further reconnections, the formation of multiple rings, 
and an energy cascade.  That is, this calculation provides,
within the context of the equations for an ideal quantum fluid,
a means by which waves on vortices can generate an energy cascade, a 
long-standing proposition for being the underlying source of the
observed decay of quantum turbulence. And not requiring filament approximations
of any type.

The results here suggest that even at higher temperatures the decay of the
quantum fluid occurs independently of the normal fluid.  To determine this
numerically would require a model that could
couple the quantum and normal fluid components of a superfluid.  Currently
a good microscopic theory of the normal fluid does not exist, so 
phenomenological models would have to suffice.  We would want a model 
where the coupling went to zero as $T\rightarrow0$ \cite{HolmX01}.

The final new analysis shows how the transformation of kinetic energy into
interaction energy is associated with: First, vortex stretching.
Then the formation of rings, followed by a signature of waves appearing in
distributions.  The formation of waves does not seem to be due to forcing
of the background quantum fluid by the vortices and their reconnections, but
instead originates from interactions between the stretched, reconnecting vortices.

\section{Summary}

The analysis of the dynamics or reconnecting anti-parallel quantum vortices has identified
the following steps: 
\ITM\item First, before vortex cores interact directly, 
Biot-Savart vortex dynamics of stretching, curvature and torsion 
dominates in a manner consistent with filament calculations and 
simulations of the classical, ideal Euler equations.
\item[] During this phase, kinetic energy is converted into interaction energy
by the vortex stretching and growth in the line length.
\item Once the cores begin to interact, the dynamics of the Gross-Piteavskii equations takes over,
with reconnection developing in the vacuum that
forms between the pair. 
\item Following the first reconnection event,
vortex waves are emitted with properties similar to waves in the local induction
approximation.  
\item These waves propagate down the initial vortex and deepen.
\item After they have deepened enough, secondary reconnections occur and vortex
rings form.  
\item Near the time of the second reconnection, 
interaction energy pulled from the original
kinetic energy is concentrated in the $\rho\approx0$ regions and
spectra have a $k^{-3}$ regime. 
\item As the vortex rings fully separate, interaction energy continues to grow,
now it appearing more on either side of $\rho=1$,
indicating either waves or rarefactions in the centers of the newly created rings.  
\item During this stage,
the high wavenumber spectra grow until spectra in two directions develop nearly
-5/3 subranges.  
\item All of these steps occur without any reduction in the global Hamiltonian. 
\ITN

The parting question posed by these calculations is to find out 
how much of this could be carried over to dynamics in the classical equations.
One point of view could be that despite the differences between the two systems, 
because the vortices in GP are well-defined and there are no singularities, this might be
a simpler system within which to identify some of the dynamics behind classical turbulence.
To answer this, new classical calculations have begun with new initial conditions for
the trajectories of vortex lines based on the calculations here.

{\bf Acknowledgments} 
We acknowledge discussions and communications with C. Barenghi, M.E. Fisher, 
W.F. Vinen, A. Golov, D.P. Lathrop, T. Lipniacki, B. Svistunov.
Partial support for this work was provided by the Leverhulme
Foundation grant F/00 215/AC. Computational support was provided 
by the Warwick Centre for Scientific Computing. 
Acknowledge support of EU-COST Aerosols and Particles

{}

\end{document}